\documentclass[nonacm,sigconf]{acmart}

\renewcommand\footnotetextcopyrightpermission[1]{}

\AtBeginDocument{%
  }

\newif\ifdraft
\draftfalse % final version

\usepackage{graphicx}
\usepackage{enumitem}
\usepackage{colortbl}
\usepackage[htt]{hyphenat}

\definecolor{neonlightgreen}{rgb}{0.22, 0.7, 0.15}
\newcommand{\shadecell}[1]{\cellcolor{neonlightgreen!#1}#1}
\newcommand{\shadeunderlinecell}[1]{\cellcolor{neonlightgreen!#1}{\underline{\textbf{#1}}}}

\renewcommand{\arraystretch}{1.5} % Increase the row spacing
\usepackage{pgfplots}
\usepackage{adjustbox}
\pgfplotsset{
    tick label style={font=\footnotesize},
    legend style={font=\footnotesize},
}

\newcommand{\gh}{GitHub}

\newcommand{\gw}{GitHub workflows}

\newcommand{\eg}{{e.g.,}\xspace}
\newcommand{\ie}{{i.e.,}\xspace}
\newcommand{\daone}{$D1$\xspace}
\newcommand{\datwo}{$D2$\xspace}
\newcommand{\dathree}{$D3$\xspace}

\newcommand{\etal}{{\em et al.}\xspace}
\newcommand{\argus}{{\sc Argus}\xspace}
\newcommand{\actionlint}{{\tt actionlint}\xspace}

\usepackage{todonotes}
\usepackage{soul}

\usepackage{booktabs}
\usepackage{multirow}
\usepackage{xspace}
\usepackage{acronym}
\usepackage{multirow}
\usepackage{tcolorbox}
\definecolor{brown(web)}{rgb}{0.65, 0.16, 0.16}
\definecolor{burgundy}{rgb}{0.5, 0.0, 0.13}
\definecolor{olivegreen}{cmyk}{0.64,0,0.85,0.40}

\newcommand{\machiry}[1]{\textcolor{green}{machiry: #1}}

\acrodef{VRSP}{Virtual Remote Security Property}
\acrodefplural{VRSP}{Virtual Remote Security Properties}
\acrodef{RCD}{Remote Control Device}
\acrodef{RCDSP}{Remote Control Device Security Property}
\acrodefplural{RCDSP}{Remote Control Device Security Properties}
\acrodef{LLM}{Large Language Models}
\acrodef{CI}{Continuous Integration}

\newcommand{\code}[1]{\texttt{#1}}

\newcommand{\ciset}{{\sc CAset}\xspace}

\newcommand{\bleu}{\emph{BLEU}}
\newcommand{\acck}{\emph{Accuracy@K}}
\newcommand{\fscore}{\emph{F1-Score}}
\newcommand{\gpt}{{GPT-3.5}}
\newcommand{\codellama}{{CodeLlama}}
\newcommand{\starchat}{{StarChat}}

\definecolor{syntxbugcolor}{rgb}{1.0, 0.49, 0.0}

\newcommand{\sect}[1]{\S~\ref{#1}}
\newcommand{\fig}[1]{Figure~\ref{#1}}
\newcommand{\tbl}[1]{Table~\ref{#1}}
\newcommand{\apdx}[1]{Appendix~\ref{#1}}
\newcommand{\superemph}[1]{\ul{\emph{#1}}\xspace}
\newcommand{\lst}[1]{Listing~\ref{#1}}

\graphicspath{ {imgs/} }

\usepackage{listings}
\usepackage[frozencache,cachedir=.]{minted}
\usepackage{fontawesome5}

\usepackage{subfig}

\usepackage{makecell}
\usepackage[flushleft]{threeparttable}

\usepackage{comment}

\usepackage{algorithmic}
\usepackage{textcomp}
\usepackage{textcase}
\usepackage{multirow}
\usepackage{booktabs}

\usepackage{wrapfig}

\usepackage{pifont}

\usepackage[hang,flushmargin]{footmisc} 

\usepackage[font={small},labelfont={bf},labelsep=colon,justification=justified]{caption}

\usepackage{tikz}
\usepackage{amsmath}

\acrodef{LLM}{Large Language Model}
\acrodef{CWE}{Common Weakness Enumeration}

\usepackage{filecontents}
\usepackage{flushend}

\newcommand\encircle[1]{\microtypesetup{deactivate}\tikz[baseline=(X.base)] \node (X) [draw, shape=circle, inner sep=0em,text width=1em, text centered] {\scriptsize #1};\microtypesetup{reactivate}}
\begin{document}

%%
%% The "title" command has an optional parameter,
%% allowing the author to define a "short title" to be used in page headers.
\title[]{On the effectiveness of Large Language Models for GitHub Workflows}

\author{Xinyu Zhang}
\affiliation{
  \institution{Purdue University}
  \country{USA}
}

\author{Siddharth Muralee}
\affiliation{
  \institution{Purdue University}
  \country{USA}
}

\author{Sourag Cherupattamoolayil}
\affiliation{
  \institution{Purdue University}
  \country{USA}
}

\author{Aravind Machiry}
\affiliation{
  \institution{Purdue University}
  \country{USA}
}

\begin{abstract}
GitHub workflows or GitHub CI is a popular continuous integration platform that enables developers to automate various software engineering tasks by specifying them as workflows,~\ie{}~\code{YAML} files with a list of jobs.
However, engineering valid workflows is tedious. They are also prone to severe security issues, which can result in supply chain vulnerabilities.
Recent advancements in~\acfp{LLM} have demonstrated their effectiveness in various software development tasks.
However,~\gw{} differ from regular programs in both structure and semantics.
We perform the first comprehensive study to understand the effectiveness of~\acp{LLM} on five workflow-related tasks with different levels of prompts.
We curated a set of $\sim$400K workflows and generated prompts with varying detail.
We also fine-tuned~\acp{LLM} on~\gh{} workflow tasks.
Our evaluation of three state-of-the-art~\acp{LLM} and their fine-tuned variants revealed various interesting findings on the current effectiveness and drawbacks of~\acp{LLM}.
%CI is important and workflows play a major role in automating all aspects.
%Workflows are good target to be automatically generated --- no complex constructs, simple dependencies, etc.
%We explore how effective are LLMs for GitHub Workflows?
\end{abstract}
\vspace{-10pt}
\settopmatter{printfolios=true}
\maketitle

%-------------------------------------------------------------------------------
% Sections
%-------------------------------------------------------------------------------
\section{Introduction}

% \explain{----Introduce GitHub Actions, explain it's popularity----}
\acf{CI} or Continuous Integration and Development (CI/CD) systems~\cite{humble2010continuous} play a crucial role in modern software development practices, automating the integration and testing of code to ensure its reliability and security.
Among the plethora of \ac{CI} platforms~\cite{travicci, circleci, gitlabci}, GitHub workflows or GitHub CI~\cite{githubci} emerges as a front-runner due to its seamless integration within the GitHub ecosystem, the ability to use third-party modules (\ie Actions), and flexibility in triggering mechanisms.
Developers use~\gw{} by defining a~\emph{pipeline or workflow}, which is a~\code{YAML} file that specifies all the details (\lst{code:motivatingexample} shows an example).
%This platform requires developers to define their pipelines termed GitHub workflows, in YAML syntax.  
%GitHub also gives developers the ability to use third-party tools, known as Actions, to automate some of the most common tasks e.g., cloning repositories.
%These workflows can be configured to run on various GitHub events, including pull requests, comments and more.

However, unlike traditional code, workflows contain a unique blend of configuration and programming logic and can incorporate snippets of multiple programming languages.
% within a single YAML file.
Valenzuela-Toledo~\etal{}~\cite{githubworkflowsevolution} demonstrated that despite the popularity of GitHub workflows, the process of engineering these workflows lacks tool support, leading to a high incidence of errors during their development. 
Furthermore, developers are known to use insecure practices, leading to security vulnerabilities unique to workflows.
This underscores the complexity of generating workflows and the need for techniques that can produce syntactically valid and secure workflows.
%This underscores the complexity of generating workflows and highlights the difficulty in producing workflows that operate error-free.

% \explain{----Large Language Medels play a significant role in software development and everyone uses them---}
\acp{LLM}~\cite{gpt4, llama} are igniting a revolution in the heart of the software development realm, automating various software engineering tasks such as coding~\cite{codellama}, crafting test cases~\cite{wang2024software}, and enriching code with documentation~\cite{nam2024using}.
Companies are embracing~\acp{LLM} at an unparalleled pace~\cite{urlana2024llms}, making artificial intelligence-guided development the standard in the industry.
Significant research has been conducted to assess the effectiveness of~\acp{LLM} for code generation tasks and to delve into the security aspects~\cite{fu2023security, copilotSP} of the code they produce. 
The findings from these studies on effectiveness of ~\acp{LLM} in generating code from a given prompt and the strategies to engineer effective prompts have practical implications for software development. They provide insights into how LLMs can be harnessed effectively in real-world scenarios.

\gw{}, although similar in their intent, vary in structure, semantics, and format (\sect{subsec:githubworkflows}) from traditional code written using various programming languages.
Also, vulnerabilities in GitHub workflows differ from regular code-level vulnerabilities because of the difference in the desired security properties of~\gw{}~\cite{characterizinggithub}.
OWASP has even created a new list for the Top 10 CI/CD security risks ~\cite{OWASPtop10secci} to raise awareness of CI/CD vulnerabilities.
Previous studies~\cite{characterizinggithub, muralee2023Argus, benedetti2022automatic, githubsecexpressioninjection} have meticulously examined the security characteristics of the GitHub CI platform, enumerating potential weaknesses inherent in~\gw{}.

With the increase in their adoption, it is imperative to understand the effectiveness of~\acp{LLM} for workflow-related tasks.
Prior works have explored the effectiveness of~\acp{LLM} on software development.
However, the difference in the structure, semantics, and security properties of workflows raises concerns regarding the generalizability of observations made by prior works for workflows.

In this paper, we tackle this problem by evaluating the effectiveness of~\acp{LLM} on workflow-related tasks.
%
%Considering the widespread popularity and adoption of the GitHub CI platform, alongside the notable ease with which misconfigurations can occur and their prevalent nature, it becomes imperative to evaluate the impact of ~\acp{LLM} within this context.
%Although prior works have explored the impacts of \acp{LLM} on software development, we expand on them by examining it in the context of GitHub workflows, keeping in mind the unique characteristics of GitHub workflows and the critical nature of the vulnerabilities that emerge.
%
% ----What we want to do?----
%In this paper, we want to investigate the effectiveness of~\acp{LLM} for engineering GitHub workflows.
Specifically, we intend to evaluate the effectiveness of~\acp{LLM} for:
\begin{itemize}
\item \textbf{RQ1: Workflow Generation (\sect{subsec:rq1results}).} How effective are~\acp{LLM} in generating workflows? How secure are these generated workflows?
\item \textbf{RQ2: Defect Detection in workflows (\sect{subsec:rq2results}).} How effective are~\acp{LLM} in finding different classes of defects in workflows?
\item \textbf{RQ3: Defect Repair in workflows (\sect{subsec:rq3results}).} How effective are~\acp{LLM} in repairing defective workflows?
\end{itemize}

% ---How did we do it?----

\noindent We selected (our selection criteria in~\sect{subsec:llmselectioncriteria}) three state-of-the-art~\acp{LLM} as our subjects,~\ie{}~\gpt{}~\cite{gpt3turbo},~\codellama{}~\cite{codellama}, and~\starchat{}~\cite{starchatbeta}.
We curated a large ($\sim$400K) workflow dataset by enhancing an existing dataset (provided by~\argus{}~\cite{muralee2023Argus}) with various prompts and syntactic defects.
We created fine-tuned variants of all our~\acp{LLM} by using a small subset of our dataset.
%We evaluated both off-the-shelf and fine-tuned variants for each~\ac{LLM}.
We evaluated both off-the-shelf and fine-tuned variants for each~\ac{LLM} along different modes (\ie{} 0-shot, 1-shot, etc).
We organized each research question into various tasks (details in~\tbl{tab:methodology}).
For each task, we designed prompts with varying levels of detail and contextual information.
For a given~\ac{LLM} variant (\ie{} off-the-shelf or fine-tuned) and a task, we first perform calibration on a small subset of data,~\ie{} to identify the temperature value and prompt that performs the best.
Second, we perform the final evaluation using the best-performing configuration on the large dataset.

%Our approach to answering these research questions began by selecting three of the state-of-the-art~\acp{LLM}.
%We used a public dataset comprising GitHub workflows and associated vulnerabilities, which served to fine-tune some of the~\acp{LLM}. 
%Additionally, we experimented with several configurations to identify the optimal combinations of prompts and temperature settings that yield the best outcomes for the tasks assigned to the ~\acp{LLM}.
%Subsequently, we conducted an evaluation of these~\acp{LLM} against each research question, benchmarking their performance against state-of-the-art tools for analysis and leveraging such tools on generated workflows or patches when the~\acp{LLM} were responsible for creation tasks.

% ---How does the results look like?---
Our study revealed various interesting findings, such as,
unlike regular code generation tasks,~\acp{LLM} requires detailed prompts to generate desired workflows.
However,~\acp{LLM} have a high likelihood of producing invalid (\ie{} with syntactic errors) workflows with detailed prompts.
Also,~\acp{LLM} can produce workflows with code injection vulnerabilities.
There is a significant difference in the performance of~\acp{LLM} in detecting different types of defects.
Also, fine-tuning reduced the effectiveness of~\starchat{} for detecting syntactic errors.
Currently,~\acp{LLM} are ineffective at repairing workflow defects, eliciting the need for novel~\acp{LLM} assisted techniques.

%\sid{this paragraph not so sure about - should I explain our experiments in more detail in %the previous paragraph so that we can say oh the result is X or Y.}
%We conducted a large scale evaluation using XXX workflows. Our experiments revealed that, LLMs require highly detailed prompts to generate desired workflows. They also have a high likelihood of producing invalid workflows, producing over XX\% of total generated workflows as invalid. We also found that LLMs have a high risk of producing vulnerable workflows, with XX\% of workflow generated having atleast one security issue. Our results show that while LLMs are adept at finding security vulnerabilities, they are unable to fix these vulnerabilities.

\noindent In summary, our contributions are as follows:
\begin{itemize}
    \item A systematic evaluation of the capabilities of three state-of-the-art~\acp{LLM} to generate~\gw{} and detect, and repair different classes of defects.
    \item Various prompt engineering techniques aimed at optimizing the performance of \acp{LLM} across various tasks related to~\gw{}.
    \item Insights about the current state and limits of \acp{LLM} when applied to engineering and security of GitHub workflows.
    \item A curated set of $\sim$400K workflows with various prompts enabling future~\ac{LLM} research on~\gw{}.
\end{itemize}
Our dataset and evaluation package will be made open-source upon publication.
\section{Background and Related Work}
\label{sec:related-work}

In this section, we will provide the necessary background on GitHub workflows,~\acp{LLM}, and discuss related work.
\subsection{GitHub workflows}
\label{subsec:githubworkflows}

% \machiry{Brief back ground of GitHub workflows and why we choose it, what are they composed of and how are they written, triggers, etc}
\begin{listing}[t]
\begin{minted}[breaklines,escapeinside=&&,fontsize=\scriptsize]{yaml}
name: Deployment
on: &\encircle{1}&
  &\textcolor{syntxbugcolor}{\faBug}& # dummy_pull:
  pull_request:
    branches: [main]
jobs: &\encircle{2}&
  build: &\encircle{8}&
    steps:
     - name: Checkout Code
       uses: actions/checkout@v2 &\encircle{3}&
     - name: Build the code
       run: make &\encircle{4}&
    ...
  test: &\encircle{9}&
    needs: build &\encircle{5}&
    steps:
     - name: Log tests
       run: |
            echo "Running tests" 
            echo "Commit - ${{ github.event.pull_request.head.sha }}" &\encircle{6}&
            &\textcolor{red}{\faBug}& echo "Branch - ${{ github.event.pull_request.head.ref }}" &\encircle{7}&
            &\textcolor{green}{\faCheck}&  # echo "Branch - $BRANCH_NAME"
       # env: 
           # BRANCH_NAME: ${{ github.event.pull_request.head.ref }} 
      ...
\end{minted}
\caption{Example of a workflow file which is triggered upon the creation of a pull request. The workflow builds the submitted code and runs the existing test-suite.}
\label{code:motivatingexample}
\end{listing}

GitHub workflows can be created by adding a~\code{YAML} file (\ie{} workflow file) to the \texttt{.github/workflows} folder in the target GitHub repository. 
The~\lst{code:motivatingexample} shows an example of a workflow file with markings representing different components.
A workflow file needs to have event triggers (which trigger the execution of workflow,~\ie{}~\encircle{1} in~\lst{code:motivatingexample}), jobs (\encircle{2}) to be executed (\eg{}~\encircle{8},~\encircle{9}), where each job is a sequence of steps (\eg{}~\encircle{3},~\encircle{4}). A job can be dependent on other jobs,~\eg{} the job~\code{test} depends on~\code{build} as indicated by~\encircle{5}.
Each step represents a unit of work, which can be performed either through running shell commands (\eg{}~\encircle{6},~\encircle{7}), programs (\eg{}~\encircle{4}), invoking other modules (\ie{} Actions)~\eg{}~\encircle{3}.
A workflow starts execution when one of the triggers occurs. Each job is independent, and all jobs execute in parallel unless there is a dependency where a job waits for all its dependents.
Within each job, steps are executed sequentially in the order specified in the workflow.

Several works studied~\gw{} along various aspects, such as most common automation practices~\cite{githubactionsdev}, common patterns to perform various tasks~\cite{devsusegithubactions}, and changes made by developers over time~\cite{githubworkflowsevolution}.
Valenzuela-Toledo~\etal{}~\cite{githubworkflowsevolution} highlighted the absence of robust tools that could support \gh{} workflows and detect syntactic and functional errors at an early stage in the development process. 
None of these works involve \acp{LLM} or have specifically addressed their use for~\gh{} workflows.

\subsubsection{Defects in workflows}
\label{subsubsec:defectsinworkflows}
Similar to traditional programs, workflows can also have defects. We focus on two classes of defects:~\emph{syntactic errors} and~\emph{security vulnerabilities}, specifically code injection vulnerabilities.

\noindent\emph{Syntactic errors} prevent the workflow from being executed.
However, identifying syntactic errors in workflows requires complete knowledge of workflow structure and valid values.
The mere validity of~\code{YAML} file does not guarantee correct workflow syntax. For instance, in the workflow in~\lst{code:motivatingexample}, changing the trigger (\ie{}~\code{pull\_request}) to an invalid name (say~\code{dummy\_pull} as indicated by~\textcolor{syntxbugcolor}{\faBug}) produces a valid~\code{YAML} but syntactically invalid workflow.

\noindent\emph{Security vulnerabilities} could be exploited by attackers to perform various malicious activities (\eg{} exfiltrating repository secrets), leveraging the permissions assigned to the workflow causing broader supply chain attacks~\cite{GithubSupplychainAttack1, GithubSupplychainAttack2}.
%Furthermore, vulnerabilities in workflows have the potential to be exploited as components of broader supply chain attacks, a risk that has been documented in existing research ~\cite{GithubSupplychainAttack1, GithubSupplychainAttack2}. 
For instance, in \lst{code:motivatingexample}, one of the steps (indicated by~\encircle{7}) prints the source branch name of a pull request and is prone to code injection vulnerability (indicated by~\textcolor{red}{\faBug}).
Note that the branch name (\code{github.event.pull\_request.head.ref}) is determined by the creator of the pull request rather than the repository owner. 
An attacker can craft a branch name that includes the desired shell command and raise a pull request.
The print command will interpret the branch name as a shell command and execute the attacker-provided command.
The~\textcolor{green}{\faCheck} marker shows the correct way to print,~\ie{} using an intermediate environment variable for the branch.

Security aspects of the GitHub CI platform have also been explored during prior research~\cite{characterizinggithub, continiousintrusion, muralee2023Argus}. 
These works primarily focus on designing static analysis tools to detect different classes of security vulnerabilities in~\gw{}.
For instance, Muralee~\etal{}~\cite{muralee2023Argus} developed~\argus{}, a static taint tracking tool aimed at identifying command injection vulnerabilities in \gh{} workflows
%Koishybayev \etal{} \cite{characterizinggithub}, for instance, characterized the security properties of \gh{} workflows. 
%They also developed GWchecker, a tool designed to identify the presence of secrets in plaintext, insecure triggers, and the use of non-updated actions. 
%More recently, Gu \etal{} \cite{continiousintrusion} introduced CIInspector, a tool that enables users to investigate potential vulnerabilities in various CI platforms including \gh{} 
% CI. 
%Lastly, Muralee \etal{} \cite{muralee2023Argus}, devised ARGUS, a static taint tracking tool aimed at identifying command injection vulnerabilities in \gh{} workflows, conducting a comprehensive analysis of over 2.8 million workflows.
However, no work tries to use~\acp{LLM} for security tasks in~\gw{}.
%tools to date have been applied to \gh{} workflows generated from or through LLMs.

%
% LLM
% 

\subsection{\acfp{LLM}}
\label{subsec:backgroundllm}
\acp{LLM} have emerged as transformative tools capable of understanding and generating human-like text based on vast amounts of data they have been trained on. 
%These models seem to be good at performing various software development tasks, such as code generation and completion, automated program repair, and even documentation. 
%They operate by analyzing the context provided in developer prompts, leveraging their extensive training on diverse coding languages and frameworks to produce accurate and relevant responses.
To elicit better responses from ~\acp{LLM}, various strategies have been formulated.
Among these, \textbf{instruction fine-tuning}~\cite{wei2022finetuned, pmlr-v202-longpre23a, peng2023instruction, xu2023mentalllm} stands out as a notable approach. 
This method involves augmenting existing pre-trained models by further training them on smaller, domain-specific, and multi-task datasets and providing detailed instructions. 
%The objective of this fine-tuning process is to refine the model's capabilities, enabling it to excel in particular domains or tasks.
%Despite these improvements, it's important to acknowledge that when these fine-tuned models encounter completely new, unseen tasks, there remains a possibility that they might not perform as expected or fail to deliver the desired outputs.
Another effective strategy to elicit better responses involves the engineering of more refined prompts~\cite{ekin2023prompt},~\ie{}~\emph{prompt engineering}, provided to the models. 
%Given the significant impact of prompting on a~\ac{LLM}'s comprehension of its designated task, improving the quality and structure of prompts often leads to improved outcomes.
The usage of~\acp{LLM} can be broadly classified into the following three modes~\cite{gpt3} based on the amount of task-specific information provided:
%Based on the amount of information provided to the \ac{LLM}, these can be broadly classified into three categories:
\begin{itemize}[leftmargin=*]
    \item \textbf{Zero-shot mode} involves presenting the~\ac{LLM} with no task specific information. The expectation is that the model, leveraging its extensive pre-training, will generate relevant outputs for entirely novel problems.
    \item \textbf{One-shot mode:} Here, we provide a single example of the prompt and the desired outcome.
    %included in the prompt to illustrate the desired outcome from the \ac{LLM}.
    The example serves to guide the model's response by providing a context or template for the task at hand.
    \item  \textbf{Few-shot mode} extends the concept of one-shot mode by providing multiple labeled examples.
\end{itemize}
It is crucial to understand that the above prompting strategies are \textbf{Tuning-free prompting}~\cite{promptingdef},~\ie{} we do not change the parameters of the pre-trained~\acp{LLM}.
%~\acp{LLM} merely generate the answers based on them.

%\xinyu{Note: In some papers, few-shot learning is to use a very small number of examples to \textbf{train} the model. Here, we cite gpt-3 paper which provides a clear definition that few-shot learning is without parameter update. }

% \machiry{Discuss works that try to assess the potential of LLMs in aiding in GitHub workflow tasks.}

% \machiry{Talk about zero-shot, one-shot and few-shot. Also, why we need only zero-shot on fine-tuned models.}

\subsubsection{Using LLMs for Automated Code Generation}

% Due to their potential in generating code automatically, several studies have targeted to create fine-tuned \ac{LLM} that are trained exclusively on code.
% Popular \acp{LLM} trained specifically to generate code include CodeLlama\cite{codellama}, CodeGen\cite{nijkamp2023codegen}, StarCoder\cite{starchatbeta}, codeT5\cite{wang-etal-2021-codet5} and Codex \cite{chen2021evaluating}.

Driven by the effectiveness of~\acp{LLM}, there has been significant interest in designing~\acp{LLM} for code-related tasks.
For instance, close-source GPT-3.5~\cite{gpt3turbo} and GPT-4~\cite{gpt4}, inheriting the capabilities of Codex~\cite{chen2021evaluating} designed specifically for programming tasks, have been extensively utilized. Other open-source code~\acp{LLM} including CodeT5\cite{wang-etal-2021-codet5}, CodeGen\cite{nijkamp2023codegen}, StarCoder\cite{starchatbeta}, CodeLlama\cite{codellama}, etc, have been successively introduced, and have demonstrated remarkable performance in software development tasks.

One of the important tasks is the text-to-code generation (\ie{} generating code based on the natural language description). However, most works focus on programming languages such as \code{Java}, \code{C/C++}, and \code{Python}.
As mentioned in~\sect{subsec:githubworkflows},~\gw{} are engineered in~\code{YAML} files.
Only few works~\cite{10247987, nijkamp2023codegen} focus on using~\acp{LLM} for generating~\code{YAML} files.
Pujar~\etal{}\cite{10247987} fine-tuned the CodeGen\cite{nijkamp2023codegen} \ac{LLM}, and evaluated its performance in generating YAML scripts for Ansible. 
Although~\gw{} follow the~\code{YAML} syntax, they differ significantly from Ansible scripts (\sect{subsec:githubworkflows}).

\subsubsection{Using LLMs for Automated Defect Detection}
Many works investigated the effectiveness of~\acp{LLM} in defect detection in regular programs.
Thapa~\etal{} \cite{TransformerModel} fine-tuned various transformer-based language models (\eg{}~BERT~\cite{bert}, GPT-2~\cite{gpt2}, DistilBERT~\cite{distilbert}, etc) on binary and multi-classification tasks using software vulnerability datasets from C/C++ applications.
Similarly, Gao~\etal{}~\cite{gao2023far} evaluated defect detection capabilities in CTF (Capture-the-Flag) challenges and real-world applications.
Fu \etal{} \cite{fu2022linevul} introduced LineVul, a line-level vulnerability predictor leveraging BERT to predict the presence of vulnerabilities in a dataset composed of C/C++ applications. 

All these works focus on vulnerabilities in regular programs.
However, as we explained in~\sect{subsubsec:defectsinworkflows}, defects in~\gw{} differ from those in regular programs.
Furthermore, none of the existing works try to evaluate the accuracy of the detection,~\ie{} line number of the defect.
In this work, we focus on holistically assessing~\acp{LLM} capabilities to detect workflow defects.

%A multitude of studies have concentrated on identifying security vulnerabilities by utilizing language models, although these studies did not specifically focus on \gh{} workflows. 
%Thapa \etal{} \cite{TransformerModel} fine-tuned various transformer-based language models (\eg{}~BERT~\cite{bert}, GPT-2~\cite{gpt2}, DistilBERT~\cite{distilbert}, etc) on binary and multi-classification tasks using software vulnerability datasets from C/C++ applications. 
%Their findings showed that these models achieved significantly superior F-1 scores compared to other contemporary models. 
%In another effort, Fu \etal{} \cite{fu2022linevul} introduced LineVul, a line-level vulnerability predictor leveraging BERT to predict the presence of vulnerabilities in a dataset composed of C/C++ applications. 
%As models continue to increase in size, researchers are shifting their focus toward larger models.
%Gao \etal{} \cite{gao2023far} evaluated~\acp{LLM}'s abilities to detect vulnerabilities in CTF(Capture-the-Flag) challenges and real-world applications, and then compared performance with that of deep learning-based models and static analyzers.

%Nevertheless, all these works concentrate on binary or multi-classification detection tasks, \ie{}, whether a code snippet contains vulnerabilities or not. They don't expect~\acp{LLM} to pinpoint the location of defects in code snippets nor analyze the defects.

\subsubsection{Using LLMs for Automated Program Repair (APR)}
Sobania \etal{} \cite{sobania2023analysis} performed a comparative evaluation of Python program repair effectiveness of ChatGPT's~\cite{gpt3turbo} with Codex and CoCoNut~\cite{coconut}.
Ahmad~\etal{}~\cite{ahmad2023fixing} employed an ensemble of~\acp{LLM}, specifically Codex and CodeGen, to automatically rectify hardware security vulnerabilities in Verilog.
Wu~\etal{}~\cite{javavulnerability} studied the capabilities of~\acp{LLM} in Java vulnerability repair and compared them with those of deep-learning-based APR models.
However, no studies focus on CI/CD platforms, specifically~\gw{}, which contain a blend of configuration steps (potentially) involving various programming languages.

\section{Study Design}
\begin{figure}[t]
    %\centering
    \includegraphics[scale=0.65]{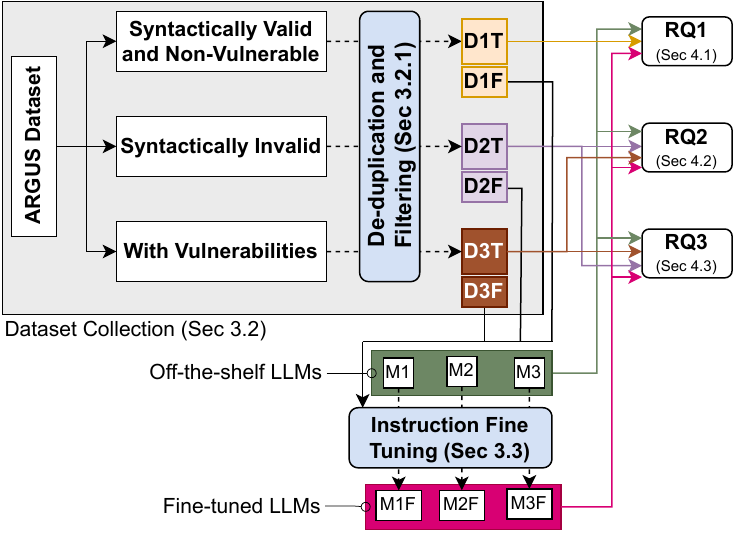}
 % https://drive.google.com/file/d/1f-_QlxK-0OUfp9Uegrw3cS6KkcNbhyHo/view?usp=sharing
    \caption{Overview of Our Study.}
    \label{fig:system_diagram}
\end{figure}

The~\fig{fig:system_diagram} shows the overview of our study. 
% We created three~\gw{} datasets,~\daone{},~\datwo{}, and~\dathree{} to investigate~\rqone{},~\rqtwo{}, and~\rqthree{}, respectively.
We created three~\gh{} workflow datasets,~\daone{},~\datwo{}, and~\dathree{} to investigate our research questions. 
We selected three state-of-the-art~\acp{LLM} and fine-tuned them with a mixed subset of our datasets.
% We selected three state-of-the-art~\acp{LLM} and fine-tuned them with our datasets.
We performed our investigation with both off-the-shelf~\acp{LLM} and their fine-tuned versions.
%We will present the details of~\acp{LLM} selection and dataset collection in this section and the details of research questions in their corresponding sections.

\subsection{\acp{LLM} Selection}
\label{subsec:llmselectioncriteria}

% Background on~\acp{LLM} and what are the possible~\acp{LLM} we use in this work?
% --May be have a table.
% Lets focus on code Generation LLMs (single shot and interactive) and ChatGPT:
% Find open-source code generation LLMs, and figure out how to interact with them.
% Original LLMs and Fine-tuned LLMs.

%-------------------------------------------------------------------------------
% In Natural language Processing(NLP),  transfomer-based\cite{transformer} models   are variants of the Transformer architecture, which excels at handling sequential data, and fall into three categories: (i) encoder-only models such as BERT\cite{bert} and RoBERTa\cite{liu2019roberta}, (ii) decoder-only models like GPT-x models \cite{gpt}\cite{gpt2}\cite{gpt3} and (iii) encoder-decoder models such as T5\cite{T5} and BART\cite{lewis-etal-2020-bart}. LLMs have recently demonstrated impressive capabilities in a wide spectrum of programming language, including code completion, code generation, code summarization, etc. All the code LLMs aim at general programming language like Java, Python and Go, but domain specific languages like YAML have received less attention. Therefore, our work try to evaluate LLMs' abilities of addressing tasks related to Github Workflows which are defined in YAML.

We aim to select state-of-the-art~\acp{LLM} that are specifically designed for programming tasks (\eg{} code completion, code generation, defect detection, etc).
We focus on instruction-following~\acp{LLM},~\ie{} which perform a task based on provided instructions.
Finally, we also should be able to fine-tune the models,~\eg{} we exclude GPT-4~\cite{gpt4} as we have no access to fine-tune it. Based on the above criteria, we selected three~\acp{LLM},~\ie{} GPT-3.5 Turbo~\cite{gpt3turbo}, StarChat-$\beta$~\cite{starchatbeta}, and CodeLlama-7B-Instruct~\cite{codellama} as summarized in~\tbl{tab:model}. 

%In this paper, we investigate three state-of-the-art pre-trained Large language models (LLMs), namely, GPT-3.5 Turbo, StarChat-$\beta$, and CodeLlama-7b-Instruct. Our choice is guided by a selection criteria focused on three primary considerations. Firstly, our preference goes to pre-trained models that have been trained on a large programming language corpus.  Secondly, we prefer pre-trained models that have been trained on instruction datasets, equipping them with the requisite prowess to comprehend the tasks based on provided instructions. For instance, StarChat is built on top of StarCoder and trained to act as coding assistants, and we choose instruction-following model within Code Llama family. Lastly, our selection leans toward pre-trained models that offer the capability for fine-tuning, \eg{} we exclude GPT-4 as we have no access to fine-tune it. The details for the models are depicted in Table \ref{tab:model}.

% Please add the following required packages to your document preamble:
% \usepackage{booktabs}
\begin{table}[h]
\caption{Models considered in our studies}
\vspace{-9pt}
\label{tab:model}
\resizebox{\columnwidth}{!}{%    
% \begin{tabular}{@{}c|c|c|c|c@{}}
% \toprule
% \textbf{ID} & \textbf{Model} & \textbf{Parameters} & \textbf{Context Length} & \textbf{Provider} \\ \midrule
% \multicolumn{1}{c|}{\textbf{M1}} & \multicolumn{1}{c|}{\textbf{GPT-3.5 Turbo}} & \multicolumn{1}{c|}{175B}  & \multicolumn{1}{c|}{4,096 tokens} & OpenAI            \\ \midrule
% \multicolumn{1}{c|}{\textbf{M2}} & \multicolumn{1}{c|}{\textbf{StarChat-$\beta$}}     & \multicolumn{1}{c|}{16B} & \multicolumn{1}{c|}{8,192 tokens} & Hugging Face      \\ \midrule
% \multicolumn{1}{c|}{\textbf{M3}} & \multicolumn{1}{c|}{\textbf{CodeLlama-7b-Instruct}}       & \multicolumn{1}{c|}{7B}   & \multicolumn{1}{c|}{16,384 tokens}             & Meta        \\ \midrule
% \end{tabular}
\begin{tabular}{@{}c|c|c|c|c@{}}
\toprule
\textbf{ID} & \textbf{Model} & \textbf{Parameters} & \textbf{Context Length} & \textbf{Provider} \\ \midrule
\textbf{M1} & ~\gpt~(GPT-3.5 Turbo) & - & 4,096 tokens & OpenAI \\ \midrule
\textbf{M2} & ~\starchat~(StarChat-$\beta$) & 16B & 8,192 tokens & Hugging Face \\ \midrule
\textbf{M3} & ~\codellama~(CodeLlama-7B-Instruct) & 7B & 16,384 tokens & Meta \\ \bottomrule
\end{tabular}

}
\end{table}

% Context length: Context length is the number of tokens a language model can process at once. It is the maximum length of the input sequence.

% max\_length (class transformers.GenerationConfig): The maximum length the generated tokens can have. Corresponds to the length of the input prompt + max\_new\_tokens. Its effect is overridden by max\_new\_tokens, if also set.

% max\_new\_tokens (class transformers.GenerationConfig): The maximum numbers of tokens to generate, ignoring the number of tokens in the prompt.

% max\_tokens (OpenAI API reference): The maximum number of tokens to generate in the completion. The token count of your prompt plus max\_tokens cannot exceed the model's context length. 

\subsection{Dataset Collection}
We used the~\gh{} workflow dataset from~\argus{}~\cite{muralee2023Argus}, a recent work that tries to find vulnerabilities in~\gw{}.
The dataset has 2,778,483 \gw{}, collated over a period from November to December 2022.
The dataset also includes 7,640~\gw{} with manually confirmed vulnerabilities.
We split the dataset into three mutually exclusive sets:
\begin{itemize}[leftmargin=*]
\item\textbf{Dataset II (D2):} This set contains an equal number of~\gw{} with one syntax error and workflows with no syntax errors.
Specifically, we ran~\actionlint{}~\cite{actionlint}, a syntax checker tool to find workflows with syntax errors, and picked the same number of syntactically valid workflows to create~\datwo{}.
\item\textbf{Dataset III (D3):} Similarly, this set contains an equal number of~\gw{} with at least one vulnerability and workflows with no vulnerabilities.
We used the vulnerable workflows from~\argus{} dataset and collected the same number of non-vulnerable workflows to create~\dathree{}.
\item\textbf{Dataset I (D1):} All the remaining workflows,~\ie{} syntactically valid and contain no vulnerabilities, are collected to form~\daone{}.
\end{itemize}

\subsubsection{De-duplication and Filtering}
We deduplicated~\argus{} dataset and ignored workflows with more than 1,024 tokens (2,048 tokens for vulnerable workflows) considering the context length supported by the selected~\acp{LLM} and our designed prompts.
%\revision{\gw{} are processed through multiple stages before splitting into three datasets. We deduplicated~\argus{} dataset, and ignored workflows with more than 1,024 tokens (2,048 tokens for vulnerable workflows) considering the context length supported by the selected~\acp{LLM} and our designed prompts. \st{We deduplicated each of our datasets and also ignored Workflows with more than 2,048 tokens,~\ie{} the maximum length supported by the selected LLMs.} }
Also, for~\daone{}, we performed the following two additional filtering steps to ensure that it contains mostly representative and realistic workflows.
First, we ignored workflows that lack names or have steps without names.
As we will discuss in~\sect{subsubsec:prompteng}, these names are needed to create prompts and are important to understand the objectives of workflows.
Second, we classified workflows using structural complexity metrics and filtered out outliers as they are not representative of realistic workflows. We provide the details of this in~\apdx{apdx:workflowcomplex}. 

\subsubsection{Fine-Tuning Dataset}
\label{subsubsec:finetuningsplit}
%\revision{\st{We also created a fine-tuning split for each dataset by randomly picking 3,200 workflows from the corresponding dataset.}
%We also created a fine-tuning split for each dataset by randomly picking the same number (3,200) of workflows from the corresponding dataset. Balancing the distribution of different tasks prevents the fine-tuned model from becoming biased towards one task. 
%}
We also created a fine-tuning split for each dataset by randomly picking the same number (3,200) of workflows from the corresponding dataset.
We capped at 3,200 as we did not find any significant increase in effectiveness with a larger number of workflows.
For~\datwo{} and~\dathree{}, we picked 1,600 positive and negative workflows.
As we will discuss in~\sect{sec:results}, we used the fine-tuning split for each dataset to create fine-tuned~\acp{LLM}.

The~\tbl{tab:dataset} shows the summary of different datasets and statistics of the corresponding workflows.

%We curate our datasets based on the ARGUS \cite{muralee2023Argus} dataset, which consists of 2,778,483 \gw{}, collated over a period from November to December 2022. The authors of ARGUS stated that code injection vulnerabilities were detected in 27,465 \gw{} and they confirmed the presence of code injection vulnerabilities in 7,640 \gw{} through manual verification. 

%We create three datasets from these \gw{}. \textbf{Dataset I}, containing syntactically valid and non-vulnerable \gw{}, serves the purpose of assessing \gh{} workflow generation. \textbf{Dataset II}, comprising both syntactically valid \gw{} and syntactically invalid ones, is utilized to identify and fix syntax errors. \textbf{Dataset III}, consisting of vulnerable \gw{} along with non-vulnerable ones, is employed to detect and repair code injection vulnerability. 

% Please add the following required packages to your document preamble:
% \usepackage{multirow}
\begin{table}[h]
\scriptsize
\caption{Summary of the different datasets used in our study.}
\vspace{-9pt} 
\label{tab:dataset}
\resizebox{\columnwidth}{!}{ 
\centering
\begin{tabular}{c|c|c|c|c}
\toprule
\multicolumn{3}{c|}{\textbf{Datasets}} & \textbf{\begin{tabular}[c]{@{}c@{}}Num\\ Workflows\end{tabular}} & 
\textbf{\begin{tabular}[c]{@{}c@{}}Size (Bytes)\\ Min/Mean/Median/Max\end{tabular}} \\ \midrule
\multirow{2}{*}{D1} & \multirow{2}{*}{Dataset I}   & FT (D1F) & 3,200 &  155/1,388/1,212/4,060 \\ \cline{3-5} 
                    &                              & Test (D1T)  & 287,876 & 84/1,247/1,068/4,450 \\ \hline
\multirow{2}{*}{D2} & \multirow{2}{*}{Dataset II}  & FT (D2F) & 3,200 & 55/1,362/1,147/4,338\\ \cline{3-5} 
                    &                              & Test (D2T)  & 122,640 & 20/1,352/1,126/4,751 \\ \hline
\multirow{2}{*}{D3} & \multirow{2}{*}{Dataset III} & FT (D3F) & 3,200 & 203/2,017/1,709/8,711 \\ \cline{3-5} 
                    &                              & Test (D3T)  & 2,006 & 194/2,049/1,748/7,854 \\ \bottomrule
\end{tabular}
}

\end{table}

\subsection{Instruction Fine-Tuning}
\label{subsec:instrfinetuning}
%10232867
Several works\cite{TransformerModel, khare2023understanding} show the effectiveness of fine-tuning~\acp{LLM} and demonstrate that they perform better than original models. We also used fine-tuned models as part of our study.

Fine-tuning requires a dataset of input and expected output pairs.
Specifically, for instruction fine-tuning, we need (instruction, output) pairs,~\ie{} natural language instruction to perform a task and the expected output.
We created the fine-tuning dataset for three of our tasks,~\ie{}, Workflow Generation (T1), Syntactic Error Identification (T2), and Code Injection Vulnerability Detection (T3) by using the corresponding fine-tuning splits (\sect{subsubsec:finetuningsplit}),~\ie{} D1F, D2F, and D3F, respectively.
For each task, we use the expected user prompt (\tbl{tab:methodology}) as its instruction and the corresponding workflow (T1) or defect location (T2 and T3) as the output.
We use the suffix~\emph{F} to indicate the fine-tuned variant of the model. For instance, GPT-3.5F indicates fine-tuned variant of~\gpt{} (\tbl{tab:model}).
Note that we used three generation tasks (instead of all five tasks) for fine-tuning.
This is because generating expected output for repair tasks (T4 and T5) is tedious, especially when there can be multiple valid but semantically equivalent repairs for a given defect.
Nonetheless, as shown by the recent work~\cite{wei2022finetuned}, the fine-tuned models on generation tasks will also perform better on other related but unseen tasks~\cite{xu2023mentalllm}. 
Based on this, our fine-tuned models are expected to perform better even on unseen defect repair tasks.

\subsubsection{Implementation Details.}
We use OpenAI's APIs to fine-tune the~\gpt{}. As for~\starchat{} and~\codellama{}, we utilize the Hugging Face implementation version of the models and fine-tune each model using the PyTorch framework with the parameter-efficient fine-tuning (PEFT) method. The fine-tuning processes for~\starchat{} and~\codellama{} are executed on a single NVIDIA A100 GPU with 80GB memory and on a cluster node running CentOS 7, utilizing Slurm (Simple Linux Utility for Resource Management) as the batch scheduler for resource and job management. Each model is fine-tuned for 5 epochs.
% We used 80\% of the fine-tuning dataset (\ie{} 7,200 --- 2,400 samples from each of D1T, D2T and D3T) for training and the rest 20\% (\ie{} 2,400 --- 3*800) for testing, maintaining a train-to-test ratio of 8:2.
We mixed D1F, D2F, and D3F as the training set and randomly selected 8,00 samples from each of D1T, D2T, and D3T for testing, maintaining a train-to-test ratio of 8:2.

\subsection{Methodology}
The aim of our study is to evaluate the effectiveness of~\acp{LLM} in performing various tasks related to~\gw{}.
Our study is organized into the following three research questions:
\begin{itemize}[leftmargin=*]
\item\textbf{RQ1: Workflow Generation:} What is the effectiveness of~\acp{LLM} in generating~\gw{} (T1)? How secure and valid are the generated workflows?
\item\textbf{RQ2: Defect Detection:} How effectively can~\acp{LLM} detect defects? Both syntactic errors (T2) and code injection vulnerabilities (T3)?
\item\textbf{RQ3: Defect Repair:} What is the effectiveness of~\acp{LLM} in repairing defects? Both syntactic errors (T4) and code injection vulnerabilities (T5)?
\end{itemize}
The~\tbl{tab:methodology} summarizes tasks associated with each research question.
We followed the same methodology to investigate all our research questions.
Specifically, for each task and workflow, we provide a prompt to~\acp{LLM} and compare their outputs with the expected output using various metrics (\sect{subsubsec:evalmetrics}).
%---PROMPT ENGINEERING--
\subsubsection{Prompt Engineering}
\label{subsubsec:prompteng} 
\setlength{\tabcolsep}{4pt}

\begin{table*}[t]
%\caption{Workflow-related tasks evaluated as part of the study. The table also includes prompts (\sect{subsubsec:prompteng}) and the metrics (\sect{subsubsec:evalmetrics}) used to evaluate each task.}
\caption{
Workflow-related tasks and corresponding prompts (\sect{subsubsec:prompteng}), and metrics (\sect{subsubsec:evalmetrics}) that are evaluated as a part of the study
}
\vspace{-9pt}  % Adjusts the space after the caption
\label{tab:methodology}
\resizebox{\linewidth}{!}{% 
\centering
\footnotesize
\begin{tabular}{|>\centering p{1cm}|>\centering p{2cm}|p{0.5cm} p{1cm} p{1cm} p{14cm}| p{1.3cm}|}
\hline
\multirow{3}{*}{\textbf{\begin{tabular}[c]{@{}c@{}}Research\\ Question\end{tabular}}} & \multirow{3}{*}{\textbf{Task}}                                        & \multicolumn{4}{c|}{\textbf{Prompt Engineering (\sect{subsubsec:prompteng})}}                                                                                                                                                                                                                                                                                                                                                                                                              & \multirow{3}{*}{\textbf{\begin{tabular}[c]{@{}c@{}}Evaluation \\ Metrics\\ (\sect{subsubsec:evalmetrics})\end{tabular}}}                                                   \\ \cline{3-6}
                                            &                                                                       & \multicolumn{2}{c|}{\textbf{System Prompt (\apdx{subsubsec:sysprompt})}}                                                                                                                                            & \multicolumn{2}{c|}{\textbf{User Prompt}}                                                                                                                                                                                                                            &                                                                                                \\ \cline{3-6}
                                            &                                                                               & \multicolumn{1}{c|}{\textbf{Persona}}                   & \multicolumn{1}{c|}{\textbf{Output format}}                                        & \multicolumn{1}{c|}{\textbf{ID}} & \multicolumn{1}{c|}{\textbf{Description}}                                                                                                                                                                                         &                                                                                                \\ \hline
\multirow{5}{*}{\textbf{RQ1}}               & \multirow{5}{*}{\textbf{\begin{tabular}[c]{@{}c@{}}Workflow \\ Generation \\ (T1)\end{tabular}}}                      & \multicolumn{1}{c|}{\multirow{5}{*}{\begin{tabular}[c]{@{}c@{}}software \\ engineer\end{tabular}}} & \multicolumn{1}{c|}{\multirow{5}{*}{\textasciigrave \textasciigrave \textasciigrave yaml \textless Workflow\textgreater \textasciigrave \textasciigrave \textasciigrave}} & \multicolumn{1}{c|}{P1}          & workflow-level information and all job IDs                                                                                                                                                                                        & \multirow{5}{*}{\begin{tabular}[c]{@{}l@{}}Accuracy@K\\ BLEU score\\ Manual valid- \\ation\end{tabular}} \\ \cline{5-6}
                                            &                                                                       & \multicolumn{1}{c|}{}                   & \multicolumn{1}{c|}{}                                                                                                 & \multicolumn{1}{c|}{P2}          & workflow-level information, all job IDs and all step names                                                                                                                                                                                                               &                                                                                                \\ \cline{5-6}
                                            &                                                                       & \multicolumn{1}{c|}{}                                                     & \multicolumn{1}{c|}{}                                                              & \multicolumn{1}{c|}{P3}          & workflow-level information, all job IDs, all step names and all dependencies that can be used                                                                                                                                                                                            &                                                                                                \\ \cline{5-6}
                                            &                                                                       & \multicolumn{1}{c|}{}                                                 & \multicolumn{1}{c|}{}                                                              & \multicolumn{1}{c|}{P4}          & workflow-level information, all job IDs, job-level information and all step names                                                                                                                                                                                                        &                                                                                                \\ \cline{5-6}
                                            &                                                                       & \multicolumn{1}{c|}{}                                                   & \multicolumn{1}{c|}{}                                                              & \multicolumn{1}{c|}{P5}          & workflow-level information, all job IDs, job-level information and step-level information                                                                                                                                                                                                       &                                                                                                \\ \hline
\multirow{5}{*}{\textbf{RQ2}}               & \multirow{2}{*}{\textbf{\begin{tabular}[c]{@{}c@{}}Syntactic Error\\ Identification (T2)\end{tabular}}}         & \multicolumn{1}{c|}{\multirow{2}{*}{\begin{tabular}[c]{@{}c@{}}software \\ engineer\end{tabular}}} & \multicolumn{1}{l|}{\multirow{2}{*}{\begin{tabular}[c]{@{}l@{}}\textless{}Yes \textit{or} No\textgreater~ | line number: ... \end{tabular}}}           & \multicolumn{1}{c|}{P1}          & Is there a syntactic error in the following GitHub workflow? \textasciigrave \textasciigrave \textasciigrave yaml \textless Workflow\textgreater \textasciigrave \textasciigrave \textasciigrave                                                                                                                            & \multirow{5}{*}{\begin{tabular}[c]{@{}l@{}}Accuracy@K\\ F1 score\end{tabular}}                       \\ \cline{5-6}
                                            &                                                                                    & \multicolumn{1}{c|}{}                                   & \multicolumn{1}{c|}{}                                                              & \multicolumn{1}{c|}{P2}          & Is there \textless{}syntactic error type\textgreater~in the following GitHub workflow?  \textasciigrave \textasciigrave \textasciigrave yaml \textless Workflow\textgreater \textasciigrave \textasciigrave \textasciigrave                                                                                                 &                                                                                                \\ \cline{2-6}
                                            & \multirow{3}{*}{\textbf{\begin{tabular}[c]{@{}c@{}}Code Injection \\ Vulnerability \\ Detection (T3)\end{tabular}}}  & \multicolumn{1}{c|}{\multirow{3}{*}{\begin{tabular}[c]{@{}c@{}}security \\ engineer\end{tabular}}} & \multicolumn{1}{c|}{\multirow{2}{*}{\begin{tabular}[c]{@{}l@{}} No \textit{or}\\ Yes | line number: ... | tainted \\ variable: ... | taint source: ... \end{tabular}}}           & \multicolumn{1}{c|}{P1}          & Is there any code injection vulnerability in the following GitHub workflow? \textasciigrave \textasciigrave \textasciigrave yaml \textless Workflow\textgreater \textasciigrave \textasciigrave \textasciigrave                                                                                                          &                                                                                                \\ \cline{5-6}
                                            &                                                                       & \multicolumn{1}{c|}{}                   & \multicolumn{1}{c|}{}                                                                                          & \multicolumn{1}{c|}{P2}          & Is there any \textless{}vulnerability type\textgreater{}~in the following GitHub workflow? \textasciigrave \textasciigrave \textasciigrave yaml \textless Workflow\textgreater \textasciigrave \textasciigrave \textasciigrave                                                                                                          &                                                                                                \\ \cline{5-6}
                                            &                                                                       & \multicolumn{1}{c|}{}                                                   & \multicolumn{1}{c|}{}                                                              & \multicolumn{1}{c|}{P3}          & Is there any code injection vulnerability in the following GitHub workflow? \textless{}hint message\textgreater{}. \textasciigrave \textasciigrave \textasciigrave yaml \textless Workflow\textgreater \textasciigrave \textasciigrave \textasciigrave                                                            &                                                                                                \\ \hline
\multirow{6}{*}{\textbf{RQ3}}               & \multirow{3}{*}{\textbf{\begin{tabular}[c]{@{}c@{}}Syntactic Error \\ Fixing (T4)\end{tabular}}}                & \multicolumn{1}{c|}{\multirow{3}{*}{\begin{tabular}[c]{@{}c@{}}software \\ engineer\end{tabular}}} & \multicolumn{1}{c|}{\multirow{3}{*}{\textasciigrave \textasciigrave \textasciigrave yaml \textless Workflow\textgreater \textasciigrave \textasciigrave \textasciigrave}}                                         & \multicolumn{1}{c|}{P1}          & Please fix syntactic errors in the following GitHub workflow. \textasciigrave \textasciigrave \textasciigrave yaml \textless Workflow\textgreater \textasciigrave \textasciigrave \textasciigrave                                                                                                          & \multirow{6}{*}{\begin{tabular}[c]{@{}l@{}}Accuracy@K\end{tabular}}        \\ \cline{5-6}
                                            &                                                                       & \multicolumn{1}{c|}{}                                                   & \multicolumn{1}{c|}{}                                                              & \multicolumn{1}{c|}{P2}          & Please fix syntactic errors in the following GitHub workflow. \textless location\textgreater. \textasciigrave \textasciigrave \textasciigrave yaml \textless Workflow\textgreater \textasciigrave \textasciigrave \textasciigrave                                                                                           &                                                                                                \\ \cline{5-6}
                                            &                                                                       & \multicolumn{1}{c|}{}                   & \multicolumn{1}{c|}{}                                                                                            & \multicolumn{1}{c|}{P3}          & Please fix syntactic errors in the following GitHub workflow. \textless location\textgreater. \textless{}error message\textgreater{}. \textasciigrave \textasciigrave \textasciigrave yaml \textless Workflow\textgreater \textasciigrave \textasciigrave \textasciigrave                                                             &                                                                                                \\ \cline{2-6}
                                            &\multirow{3}{*}{\textbf{\begin{tabular}[c]{@{}c@{}}Code Injection \\ Vulnerability \\ Repair (T5)\end{tabular}}}     & \multicolumn{1}{c|}{\multirow{3}{*}{\begin{tabular}[c]{@{}c@{}}security \\ engineer\end{tabular}}} & \multicolumn{1}{c|}{\multirow{3}{*}{\textasciigrave \textasciigrave \textasciigrave yaml \textless Workflow\textgreater \textasciigrave \textasciigrave \textasciigrave}}                                         & \multicolumn{1}{c|}{P1}          & Please repair code injection vulnerabilities in the following GitHub workflow. \textasciigrave \textasciigrave \textasciigrave yaml \textless Workflow\textgreater \textasciigrave \textasciigrave \textasciigrave                                                                                                      &                                                                                                \\ \cline{5-6}
                                            &                                                                       & \multicolumn{1}{c|}{}                   & \multicolumn{1}{c|}{}                                                                                             & \multicolumn{1}{c|}{P2}          & Please repair code injection vulnerabilities in the following GitHub workflow. \textless{}location \textgreater{}. \textasciigrave \textasciigrave \textasciigrave yaml \textless Workflow\textgreater \textasciigrave \textasciigrave \textasciigrave &                                                                                                \\ \cline{5-6}
                                            &                                                                       & \multicolumn{1}{c|}{}                   & \multicolumn{1}{c|}{}                                                                                              & \multicolumn{1}{c|}{P3}          & Please repair code injection vulnerabilities in the following GitHub workflow. \textless{}location\textgreater{}. \textless{}guidance on resolution\textgreater{}. \textasciigrave \textasciigrave \textasciigrave yaml \textless Workflow\textgreater \textasciigrave \textasciigrave \textasciigrave &                                                                                                \\ \hline
\end{tabular}
}
\end{table*}
Several works~\cite{velasquez2023prompt,arvidsson2023prompt} show that prompts greatly influence the effectiveness of~\acp{LLM}.
For each task, we created prompts (mimicking~\textit{user} instructions) with varying levels of detail describing the desired output from a~\ac{LLM}.

Salewski~\etal{}~\cite{salewski2023incontext} demonstrated that assigning a specific persona (\eg{} domain expert) to~\acp{LLM} will result in better results.
Based on this, we create a persona prompt or~\textit{system} prompt for each task that sets up the desired persona of a~\ac{LLM}.
\superemph{We prepend the system prompt to the user prompt to create the final prompt, which we provide to LLMs}.
The details of the prompts are depicted in~\tbl{tab:methodology}.

\noindent\emph{User Prompts for Workflow Generation (T1).} 
In this task, we evaluate the capability of~\acp{LLM} in generating well-formatted workflows from a natural language description.
As described in~\sect{subsec:githubworkflows}, a workflow has a name, trigger, and set of jobs, each with a sequence of steps.
Each of these components has a name describing its functionality,~\eg{} \emph{``Build the project''}.
We create five types of prompts (P1-P5) for this task, each prompt providing more description about the target workflow.
P1 has the minimal description needed to create the workflow,~\ie{} name, trigger, and the set of job IDs.
However, it does not provide any details about the steps in each job.
Meanwhile, P2 (in addition to information from P1) provides information about the steps.
Similarly, P3-P5 provides an increasing level of detail.
The~\tbl{tab:prompt_eg} in the Appendix shows these prompts for the workflow in~\lst{listing:wfsgen_example_gt}.

\noindent\emph{User Prompts for Syntactic Error (T2) and Code Injection Vulnerability (T3) Identification.} 
Here, we evaluate the defect detection capability of~\acp{LLM}.
We create prompts, each providing more information about the target defect.
The corresponding rows in~\tbl{tab:methodology} provide more details.
The basic prompt asks for the existence of the desired defect (\ie{} syntactic error or code injection vulnerability).
Other prompts provide more details about the target defect,~\ie{} specific type, hint message (in~\apdx{subsubsec:promptdetails}), etc. 

\noindent\emph{User Prompts for Syntactic Error (T4) and Code Injection Vulnerability (T5) Repair.} 
Here, we evaluate the defect repair capabilities of~\acp{LLM}.
We provide varying degrees of information related to the target defect, fix strategies, and templates.
These templates range from minimal information to comprehensive hints, encompassing details such as defect locations, specific error messages, and guidance on bug resolution (in~\apdx{subsubsec:promptdetails}).

%---EVALUATION METRICS---
\subsubsection{Evaluation Metrics}
\label{subsubsec:evalmetrics}
We used the following three evaluation metrics to assess the output produced by~\acp{LLM} across various tasks.

\noindent\textbf{\textit{BLEU (Bilingual Evaluation Understudy) score}}~\cite{bleuscore} is a value ranging from 0 to 1 indicating how similar the candidate text is to the reference text, with values closer to 1 representing higher similarity.
We use BLEU-4 (\ie{} the geometric average of 1-gram, 2-gram, 3-gram, and 4-gram precision) to compare the generated workflow with the expected workflow because of the need to preserve the ordering of tokens.

\noindent\textbf{\acck}~\cite{10232867}
enables measuring accuracy of results when multiple (\ie{} $K$) responses are provided.
Specifically, given a test $t$ (or a sample $s$), we consider the responses of a~\ac{LLM} as a match (\ie{} score 1) when any one of the $K$ responses satisfies $t$ (or matches $s$), else we consider it as no match (\ie{} score 0).
For $n$ tests (or samples), we average the matching score (\ie{} 1 or 0) across all the $n$ samples to get~\acck{}.

\noindent\textbf{\fscore}~\cite{sasaki2007truth} is calculated as a harmonic mean of precision and recall. This score (ranging from 0-1) provides a single metric to evaluate binary classification. We use this to evaluate defect detection effectiveness.

The last column of~\tbl{tab:methodology} shows the summary of metrics used to evaluate each task.

\noindent\emph{Workflow Generation Task (T1):}
Here, we want to evaluate whether the workflow generated by a~\ac{LLM} performs the functionality as the expected workflow.
However, precisely accessing this requires semantic equivalence checking~\cite{necula2000translation} --- infeasible in the general case.
Instead, (i) We check that the generated workflow is valid (\ie{} without syntactic errors) by using~\acck{}; and (ii) Compare how similar (content-wise) is the generated workflow compared to the expected workflow using~\bleu{} score; and (iii) Manual validation on 270 randomly sampled workflows.

\noindent\emph{Defect Detection Tasks (T2 and T3):}
Here, we verify two aspects: detection capability and accuracy of the detection.
Specifically, we use~\fscore{} to measure detection capability and measure detection accuracy (\ie{} line number for T2, line number, tainted value and taint source for T3) using~\acck{}.

\noindent\emph{Defect Repair Tasks (T4 and T5):}
Here, we want to evaluate whether a~\ac{LLM} correctly repaired a workflow.
However, automatically checking whether~\acp{LLM} produced the correct repaired workflow requires semantic checking --- similar to the workflow generation task (T1).
Instead, we check whether the generated workflow is non-vulnerable by using~\acck{}.

%---LLMs Configuration and Experimental Setup---
\subsubsection{\acp{LLM} Configuration and Experimental Setup}
As mentioned in~\sect{subsec:backgroundllm}, there are three basic modes (\ie{} zero-shot, one-shot, and few-shot) of using a~\ac{LLM} model. 
However, during our experiments with off-the-shelf variants, we found no difference in effectiveness between the one-shot mode and the few-shot mode.
\superemph{We will only present the results of zero-shot and one-shot modes for off-the-shelf variants.
For Defect Repair Tasks, we used both zero-shot mode and one-shot mode for fine-tuned variants, as these tasks are unseen to the fine-tuned models.}~(\sect{subsec:backgroundllm}).
For a given mode, the performance of a~\ac{LLM} model might vary with different configurations.
For every task, we want to assess an~\ac{LLM} mode using its best-performing configuration and the most effective prompt.

\acp{LLM} have a temperature parameter, indicating the desired level of randomness. Specifically, higher temperature values indicate a higher degree of non-determinism.
The values from 0 to 1 are recommended to prompt a~\ac{LLM} to produce responses that are acceptable to humans. For example, the temperature range of~\gpt{} is from 0 to 2. However, temperature values above 0.9 make the responses technically useless.
Also, as mentioned in~\sect{subsubsec:prompteng}, we generate several prompts for each task.

\noindent\emph{Calibration (Identifying Effective Configuration):} For a given~\ac{LLM} and task, we use a small but representative subset (\ie{} calibration set (\ciset{})) of samples to identify which temperature and prompt combination gives the best result.
Specifically, we use 0.1, 0.3, 0.5, 0.7, and 0.9 as our temperature values and combine them with prompts with varying levels of detail (\tbl{tab:methodology}).
\superemph{The best-performing temperature value and prompt will be used to evaluate the final set of samples.}

For T1, we collected 266 workflows as our~\ciset{}.
We performed a random sampling and collected two workflows each for 133 effective combinations of complexity metrics (\apdx{apdx:workflowcomplex}), ensuring that our~\ciset{} is representative.
Also, given the large number (0.28 million) of workflows in D1T, we picked 20 workflows along each of the 133 complexity metrics combinations as our evaluation dataset.

For T2, we randomly selected 200 workflows (100 syntactically valid, 100 syntactically invalid) to construct our~\ciset{} and sampled 5,000~\gw{} (2,500 syntactically valid, 2,500 syntactically invalid) to form a larger evaluation dataset.
Similarly, for T3, we randomly selected 80 vulnerable workflows (with a total of 108 vulnerabilities), and then sampled 80 non-vulnerable~\gw{} to construct~\ciset{}. We utilize all remaining vulnerable~\gw{} (923~\gw{} with 1,586 vulnerabilities) and 923 non-vulnerable~\gw{} to form a larger evaluation dataset.

For T4, we randomly chose 200~\gw{} with syntactic errors as our~\ciset{} and sampled an additional syntactically invalid 2,500~\gw{} to constitute a larger evaluation dataset. For T5, we randomly selected 100 and 375~\gw{} containing code injection vulnerabilities that can be fixed within workflows to form our~\ciset{} and larger evaluation dataset, respectively.

\section{Results}
\label{sec:results}
In this section, we present the results of the study along with our three research questions.
For each task (under a research question), we first present the calibration results,~\ie{} the most effective configuration for each~\ac{LLM} in zero-shot and one-shot modes for off-the-shelf variant and zero-shot (and one-shot for unseen tasks~\ie{} T4 and T5) for fine-tuned variant.
Second, we present the final assessment of each~\ac{LLM} mode using its most effective configuration.

\subsection{RQ1: Workflow Generation}
\label{subsec:rq1results}
Here, we evaluate the effectiveness of~\acp{LLM} in generating workflows.
As shown in~\tbl{tab:methodology} and described in~\sect{subsubsec:evalmetrics}, this research question has one task, and we use three evaluation metrics.
\subsubsection{Calibration}
We use~\bleu{} (\tbl{tab:wfsgen_step1_bleu}) and~\acck{} (\tbl{tab:wfsgen_step1_static_check_accuracy}) scores for calibration.
{
\renewcommand{\arraystretch}{1.1} % Increase the row spacing
\setlength{\tabcolsep}{6pt} % Increase the column spacing
\begin{table}[h]
    \centering
    \caption{BLEU scores of workflow generation on~\ciset{}.}
    \vspace{-9pt}
\resizebox{\columnwidth}{!}{% 
    \begin{tabular}{|c|*{16}{c|}}
    \hline
    \multirow{3}{*}{\textbf{Model}} & \multirow{3}{*}{\textbf{t}} & \multicolumn{10}{c|}{\textbf{off-the-shelf}} & \multicolumn{5}{c|}{\multirow{2}{*}{\textbf{fine-tuned}}} \\
    \cline{3-12}
    & & \multicolumn{5}{c|}{\textbf{0-shot}} & \multicolumn{5}{c|}{\textbf{1-shot}} & \multicolumn{5}{c|}{} \\
    \cline{3-17}
    & & P1 & P2 & P3 & P4 & P5 & P1 & P2 & P3 & P4 & P5 & P1 & P2 & P3 & P4 & P5 \\ \hline
    \multirow{5}{*}{\textbf{\gpt{}}}
    & 0.1 & \shadecell{13.60} & \shadecell{35.57} & \shadecell{40.87} & \shadecell{40.56} & \shadecell{74.79} & \shadecell{25.69} & \shadecell{46.20} & \shadecell{54.31} & \shadecell{54.26} & \shadecell{78.99} & \shadecell{25.83} & \shadecell{47.25} & \shadecell{53.64} & \shadecell{53.33} & \shadecell{80.84} \\
    & 0.3 & \shadecell{15.21} & \shadecell{36.92} & \shadecell{42.10} & \shadecell{41.63} & \shadecell{76.00} & \shadecell{27.77} & \shadecell{47.76} & \shadecell{55.90} & \shadecell{55.87} & \shadecell{79.71} & \shadecell{27.51} & \shadecell{49.05} & \shadecell{54.43} & \shadecell{54.72} & \shadecell{82.56} \\
    & 0.5 & \shadecell{16.00} & \shadecell{38.04} & \shadecell{43.00} & \shadecell{42.58} & \shadecell{76.98} & \shadecell{28.84} & \shadecell{48.57} & \shadecell{56.61} & \shadecell{56.47} & \shadecell{80.49} & \shadecell{27.46} & \shadecell{49.05} & \shadecell{54.60} & \shadecell{55.04} & \shadecell{83.14} \\
    & 0.7 & \shadecell{16.39} & \shadecell{38.08} & \shadecell{43.35} & \shadecell{43.17} & \shadecell{76.93} & \shadecell{28.95} & \shadecell{49.49} & \shadecell{57.23} & \shadecell{57.22} & \shadecell{81.24} & \shadecell{26.69} & \shadecell{47.79} & \shadecell{53.75} & \shadecell{55.13} & \shadecell{82.97} \\
    & 0.9 & \shadecell{16.84} & \shadecell{38.39} & \shadecell{43.69} & \shadecell{43.22} & \shadeunderlinecell{77.45} & \shadecell{29.21} & \shadecell{49.67} & \shadecell{57.15} & \shadecell{57.31} & \shadeunderlinecell{81.61} & \shadecell{24.36} & \shadecell{46.43} & \shadecell{52.54} & \shadecell{52.80} & \shadeunderlinecell{83.49} \\ \hline
    \multirow{5}{*}{\textbf{\codellama{}}} 
    & 0.1 & \shadecell{15.59} & \shadecell{35.32} & \shadecell{40.38} & \shadecell{41.98} & \shadecell{74.03} & \shadecell{36.14} & \shadecell{51.22} & \shadecell{55.81} & \shadecell{57.15} & \shadecell{79.84} & \shadecell{25.38} & \shadecell{45.54} & \shadecell{51.06} & \shadecell{53.92} & \shadecell{82.15} \\
    & 0.3 & \shadecell{17.12} & \shadecell{37.84} & \shadecell{42.56} & \shadecell{44.98} & \shadecell{75.47} & \shadecell{37.42} & \shadecell{52.68} & \shadecell{57.34} & \shadecell{58.50} & \shadecell{81.58} & \shadecell{27.43} & \shadecell{48.38} & \shadecell{53.12} & \shadecell{56.29} & \shadecell{83.73} \\
    & 0.5 & \shadecell{17.91} & \shadecell{37.56} & \shadecell{43.15} & \shadecell{45.35} & \shadeunderlinecell{76.41} & \shadecell{37.77} & \shadecell{53.20} & \shadecell{57.81} & \shadecell{59.48} & \shadeunderlinecell{82.53} & \shadecell{27.42} & \shadecell{48.60} & \shadecell{52.90} & \shadecell{56.54} & \shadeunderlinecell{84.11} \\
    & 0.7 & \shadecell{17.65} & \shadecell{37.61} & \shadecell{42.65} & \shadecell{44.43} & \shadecell{75.81} & \shadecell{38.42} & \shadecell{53.41} & \shadecell{57.72} & \shadecell{58.86} & \shadecell{81.75} & \shadecell{26.71} & \shadecell{47.08} & \shadecell{53.16} & \shadecell{55.58} & \shadecell{83.85} \\
    & 0.9 & \shadecell{16.43} & \shadecell{36.18} & \shadecell{40.38} & \shadecell{41.65} & \shadecell{73.31} & \shadecell{37.84} & \shadecell{51.83} & \shadecell{55.99} & \shadecell{57.64} & \shadecell{81.00}  & \shadecell{25.08} & \shadecell{45.37} & \shadecell{51.19} & \shadecell{53.53} & \shadecell{83.25}\\ \hline
    \multirow{5}{*}{\textbf{\starchat{}}} 
    & 0.1 & \shadecell{17.22} & \shadecell{36.87} & \shadecell{40.21} & \shadecell{41.07} & \shadecell{62.75} & \shadecell{34.61} & \shadecell{49.72} & \shadecell{53.91} & \shadecell{55.53} & \shadecell{75.38} & \shadecell{26.98} & \shadecell{48.72} & \shadecell{53.57} & \shadecell{54.47} & \shadecell{80.71} \\
    & 0.3 & \shadecell{18.54} & \shadecell{38.21} & \shadecell{41.82} & \shadecell{43.38} & \shadecell{64.83} & \shadecell{36.51} & \shadecell{51.38} & \shadecell{55.30} & \shadecell{57.36} & \shadecell{76.89} & \shadecell{29.07} & \shadecell{50.67} & \shadecell{55.75} & \shadecell{57.43} & \shadecell{81.81} \\
    & 0.5 & \shadecell{19.46} & \shadecell{39.10} & \shadecell{42.93} & \shadecell{43.81} & \shadecell{66.26} & \shadecell{37.36} & \shadecell{52.33} & \shadecell{56.58} & \shadecell{58.10} & \shadecell{77.49} & \shadecell{29.45} & \shadecell{50.74} & \shadecell{56.20} & \shadecell{58.39} & \shadeunderlinecell{82.79} \\
    & 0.7 & \shadecell{19.34} & \shadecell{39.04} & \shadecell{42.69} & \shadecell{43.81} & \shadeunderlinecell{66.45} & \shadecell{37.36} & \shadecell{52.30} & \shadecell{55.86} & \shadecell{58.40} & \shadeunderlinecell{77.88} & \shadecell{28.72} & \shadecell{50.12} & \shadecell{55.42} & \shadecell{57.15} & \shadecell{82.37} \\
    & 0.9 & \shadecell{18.91} & \shadecell{39.03} & \shadecell{42.46} & \shadecell{43.68} & \shadecell{66.31} & \shadecell{37.14} & \shadecell{51.98} & \shadecell{56.27} & \shadecell{58.00} & \shadecell{77.48} & \shadecell{26.09} & \shadecell{47.88} & \shadecell{53.35} & \shadecell{55.78} & \shadecell{82.34} \\ \hline
    \end{tabular}
    }

    \label{tab:wfsgen_step1_bleu}
\end{table}
}

% Trend of BLUE scores
\noindent\emph{\bleu{} Scores.} The~\tbl{tab:wfsgen_step1_bleu} shows that across all temperature ($t$) values and~\acp{LLM} modes.
The trend of~\bleu{} scores across different temperature values changes across different~\acp{LLM}.
For~\gpt{}, the largest temperature value of 0.9 (\ie{} greater non-determinism) is better. Whereas for~\codellama{} and~\starchat{}, temperature values of 0.5 and 0.7, respectively, are the best.
Interestingly, the~\bleu{} score increases with more detailed prompts.
This indicates that users should provide detailed prompts to get the expected workflow.
This differs from the standard code generation tasks, where~\acp{LLM} are shown to perform well even with a very simple prompt~\cite{fagadau2024analyzing}.
This is because a simple prompt can precisely describe the desired code generation task,~\eg{}~\emph{``generate sort function''}. Whereas workflows (as explained in~\sect{subsec:githubworkflows}) are sequences of steps and are hard to describe in a simple prompt. Furthermore, even for a single step, the appropriate way to perform it depends on the target project.
For instance, a step to build a project depends on the target project,~\ie{} C/C++ (make/cmake), python (\texttt{setup.py}), java (\texttt{ant build}), etc.
More contextual information is needed to generate appropriate steps and workflows.

\begin{tcolorbox} [width=\linewidth, colback=yellow!30!white, top=1pt, bottom=1pt, left=2pt, right=2pt]
\textbf{Finding 1.1:} Unlike for regular code generation tasks,~\acp{LLM} require detailed prompts to generate desired workflows.
\end{tcolorbox}

\noindent\emph{\acck{} Scores.} The~\tbl{tab:wfsgen_step1_static_check_accuracy} shows the trend of~\acck{} scores. It is interesting to see that detailed prompts do not always improve the~\acck{} scores.
In fact, detailed prompts reduce the~\acck{} scores, as shown by the decrease in scores across the P2 and P3 columns.
In other words, detailed prompts result in~\acp{LLM} producing defective workflows.

{
\renewcommand{\arraystretch}{1.1} % Increase the row spacing
\setlength{\tabcolsep}{6pt} % Increase the column spacing
\begin{table}[h]
    \caption{Accuracy@K of workflow generation on~\ciset{}.}
    \vspace{-9pt}
    \centering
\resizebox{\columnwidth}{!}{% 
    \begin{tabular}{|c|*{16}{c|}}
    \hline
    \multirow{3}{*}{\textbf{Model}} & \multirow{3}{*}{\textbf{t}} & \multicolumn{10}{c|}{\textbf{off-the-shelf}} & \multicolumn{5}{c|}{\multirow{2}{*}{\textbf{fine-tuned}}} \\
    \cline{3-12}
    & & \multicolumn{5}{c|}{\textbf{0-shot}} & \multicolumn{5}{c|}{\textbf{1-shot}} & \multicolumn{5}{c|}{} \\
    \cline{3-17}
    & & P1 & P2 & P3 & P4 & P5 & P1 & P2 & P3 & P4 & P5 & P1 & P2 & P3 & P4 & P5 \\ \hline
    \multirow{5}{*}{\textbf{\gpt{}}}
    & 0.1 & \shadecell{69.55} & \shadecell{66.17} & \shadecell{57.89} & \shadecell{60.53} & \shadecell{78.95} & \shadecell{89.10} & \shadecell{82.71} & \shadecell{70.68} & \shadecell{83.46} & \shadecell{88.35} & \shadecell{92.86} & \shadecell{87.22} & \shadecell{79.32} & \shadecell{85.71} & \shadecell{83.83} \\
    & 0.3 & \shadecell{77.07} & \shadecell{74.06} & \shadecell{62.41} & \shadecell{67.29} & \shadecell{84.21} & \shadecell{91.35} & \shadecell{84.96} & \shadecell{77.82} & \shadecell{87.22} & \shadecell{88.72} & \shadecell{95.49} & \shadecell{93.23} & \shadecell{86.09} & \shadecell{92.11} & \shadecell{89.85} \\
    & 0.5 & \shadecell{84.59} & \shadecell{78.57} & \shadecell{66.54} & \shadecell{68.80} & \shadecell{87.59} & \shadecell{90.98} & \shadecell{87.97} & \shadecell{77.07} & \shadecell{87.97} & \shadecell{89.47} & \shadeunderlinecell{96.62} & \shadecell{94.36} & \shadecell{85.71} & \shadecell{93.61} & \shadecell{88.72} \\
    & 0.7 & \shadecell{85.71} & \shadecell{82.33} & \shadecell{65.79} & \shadecell{69.92} & \shadecell{87.22} & \shadecell{92.86} & \shadecell{87.97} & \shadecell{79.70} & \shadecell{87.97} & \shadecell{92.86} & \shadecell{94.74} & \shadecell{93.98} & \shadecell{87.97} & \shadecell{93.98} & \shadecell{92.86} \\
    & 0.9 & \shadecell{89.10} & \shadecell{78.95} & \shadecell{68.42} & \shadecell{76.32} & \shadeunderlinecell{89.85} & \shadecell{91.73} & \shadecell{90.98} & \shadecell{83.46} & \shadecell{89.85} & \shadeunderlinecell{93.61} & \shadecell{90.98} & \shadecell{92.48} & \shadecell{84.96} & \shadecell{91.73} & \shadecell{92.11} \\ \hline
    \multirow{5}{*}{\textbf{\codellama{}}} 
    & 0.1 & \shadecell{84.09} & \shadecell{78.41} & \shadecell{70.83} & \shadecell{76.52} & \shadecell{73.11} & \shadecell{90.91} & \shadecell{79.55} & \shadecell{75.00} & \shadecell{77.65} & \shadecell{80.45} & \shadecell{96.21} & \shadecell{93.18} & \shadecell{88.26} & \shadecell{90.91} & \shadecell{87.12} \\
    & 0.3 & \shadecell{92.42} & \shadecell{84.85} & \shadecell{75.76} & \shadecell{83.33} & \shadecell{79.17} & \shadeunderlinecell{95.49} & \shadecell{88.35} & \shadecell{80.08} & \shadecell{83.83} & \shadecell{87.97} & \shadecell{97.73} & \shadecell{94.70} & \shadecell{92.05} & \shadecell{93.94} & \shadecell{90.91} \\
    & 0.5 & \shadecell{92.80} & \shadecell{88.64} & \shadecell{78.03} & \shadecell{85.23} & \shadecell{80.30} & \shadeunderlinecell{95.49} & \shadecell{86.47} & \shadecell{81.95} & \shadecell{85.71} & \shadecell{89.47} & \shadeunderlinecell{99.24} & \shadecell{96.21} & \shadecell{92.05} & \shadecell{96.21} & \shadecell{90.91} \\
    & 0.7 & \shadeunderlinecell{93.56} & \shadecell{88.26} & \shadecell{76.14} & \shadecell{79.92} & \shadecell{82.58} & \shadecell{93.98} & \shadecell{91.73} & \shadecell{79.70} & \shadecell{91.35} & \shadecell{86.09} & \shadecell{97.73} & \shadecell{96.97} & \shadecell{94.32} & \shadecell{96.21} & \shadecell{93.18} \\
    & 0.9 & \shadecell{83.71} & \shadecell{76.52} & \shadecell{63.64} & \shadecell{71.97} & \shadecell{75.76} & \shadecell{90.98} & \shadecell{83.83} & \shadecell{78.20} & \shadecell{81.58} & \shadecell{87.59}  & \shadecell{96.97} & \shadecell{93.94} & \shadecell{92.80} & \shadecell{92.80} & \shadecell{93.56}\\ \hline
    \multirow{5}{*}{\textbf{\starchat{}}} 
    & 0.1 & \shadecell{74.62} & \shadecell{61.74} & \shadecell{56.06} & \shadecell{57.58} & \shadecell{54.55} & \shadecell{70.08} & \shadecell{62.50} & \shadecell{59.09} & \shadecell{62.12} & \shadecell{60.53} & \shadecell{73.11} & \shadecell{70.45} & \shadecell{64.02} & \shadecell{66.29} & \shadecell{58.33} \\
    & 0.3 & \shadecell{84.85} & \shadecell{70.08} & \shadecell{62.88} & \shadecell{63.26} & \shadecell{58.33} & \shadecell{77.82} & \shadecell{71.80} & \shadecell{63.53} & \shadecell{65.79} & \shadecell{61.65} & \shadecell{83.33} & \shadecell{79.17} & \shadecell{70.45} & \shadecell{71.59} & \shadecell{59.47} \\
    & 0.5 & \shadecell{85.23} & \shadecell{74.24} & \shadecell{69.70} & \shadecell{70.83} & \shadecell{61.36} & \shadecell{78.95} & \shadecell{73.68} & \shadecell{66.92} & \shadecell{70.30} & \shadecell{63.91} & \shadecell{88.64} & \shadecell{80.30} & \shadecell{72.73} & \shadecell{76.52} & \shadecell{60.23} \\
    & 0.7 & \shadecell{87.88} & \shadecell{79.92} & \shadecell{71.97} & \shadecell{74.24} & \shadecell{62.50} & \shadecell{83.46} & \shadecell{74.06} & \shadecell{68.80} & \shadecell{72.56} & \shadecell{64.66} & \shadeunderlinecell{90.15} & \shadecell{84.85} & \shadecell{76.14} & \shadecell{78.03} & \shadecell{61.74} \\
    & 0.9 & \shadeunderlinecell{89.77} & \shadecell{76.52} & \shadecell{75.76} & \shadecell{75.00} & \shadecell{65.15} & \shadeunderlinecell{89.10} & \shadecell{74.44} & \shadecell{70.30} & \shadecell{73.31} & \shadecell{63.53} & \shadecell{83.33} & \shadecell{79.55} & \shadecell{74.62} & \shadecell{77.65} & \shadecell{64.77} \\ \hline
    \end{tabular}
    }

    \label{tab:wfsgen_step1_static_check_accuracy}
\end{table}
}

Interestingly,~\acck{} score follows a inverse bell curve for~\gpt{} and~\codellama{}.
Specifically, for low-detail prompts, the~\acck{} score decreases as the prompt becomes more detailed (till P3). However, the~\acck{} slowly rises as the prompt becomes increasingly detailed (P4 and P5).
The case is slightly different for~\starchat{}, where~\acck{} always decreases with the increase in the details of the prompt.

The trend is different for~\bleu{} score where detailed prompts provide better results.
Upon investigation, we found that~\acp{LLM} generate smaller workflows with simpler prompts and consequently reduces the chances of having defects resulting in higher~\acck{} score.
However, simpler prompts are unlikely to generate the desired workflows, as shown by the lower~\bleu{} scores (\tbl{tab:wfsgen_step1_bleu}).
On the other hand, detailed prompts to~\acp{LLM} produce workflows closer to the desired workflows, but the generated workflows might have defects. \lst{listing:wfsgen_example_gen} shows two~\gw{} generated by fine-tuned~\codellama{}. A detailed prompt (P5) produces the left workflow, which is closer to the desired workflow (\lst{listing:wfsgen_example_gt} in Appendix) but contains a syntactic error (\textcolor{syntxbugcolor}{\faBug}), while a simple prompt (P1) generates the right one which is syntactically valid but incorrect.

\begin{listing}[t]
\setlength{\tabcolsep}{0pt}{
\begin{tabular}{ll}
\begin{minipage}[t]{.25\textwidth}
\begin{minted}[breaklines, fontsize=\tiny, escapeinside=\#\#]{yaml}
name: Generate Docs
on:
  #\textcolor{syntxbugcolor}{\faBug push: \{branch: 'main'\}}#
jobs:
  build:
    runs-on: ubuntu-latest
    steps:
      - name: Checkout sources
        uses: actions/checkout@v2
      - name: Install stable toolchain
        uses: actions-rs/toolchain@v1
        with:
          profile: minimal
          target: wasm32-wasi
          toolchain: stable
          override: true
      - name: Run cargo build
        uses: actions-rs/cargo@v1
        with:
          command: doc
          args: --workspace --no-deps
      - name: Prepare docs folder
        run: |
          sudo chown -R $(whoami) target/doc
          touch target/doc/.nojekyll
      - name: Deploy documentation branch
        uses: JamesIves/github-pages-deploy-action@3.7.1
        with:
          GITHUB_TOKEN: ${{ secrets.GITHUB_TOKEN }}
          BRANCH: gh-pages
          FOLDER: target/doc
\end{minted}
\end{minipage} &
\begin{minipage}[t]{.25\textwidth}
\begin{minted}[breaklines, fontsize=\tiny]{yaml}
name: Generate Docs
on:
  push:
    branches:
      - main
jobs:
  build:
    runs-on: ubuntu-latest
    steps:
      - name: Checkout
        uses: actions/checkout@v2
      - name: Set up Rust
        uses: actions-rs/toolchain@v1
        with:
          profile: minimal
          toolchain: stable
          override: true
          components: clippy
          target: x86_64-unknown-linux-gnu
      - name: Cargo packages
        uses: actions-rs/cargo@v1
        with:
          command: build
          args: --all
      - name: Publish to gh-pages
        uses: crazy-max/ghaction-github-pages@v2
        with:
          target_branch: gh-pages
          build_dir: target/doc
\end{minted}
\end{minipage}
\end{tabular}}

\caption{Two~\gw{} generated by fine-tuned~\codellama{} with 0-shot prompting at temperature 0.9. The left workflow was generated using P5, whereas the right one was generated by P1. The expected (i.e., ground truth) workflow is in~\lst{listing:wfsgen_example_gt} (Appendix).}
\label{listing:wfsgen_example_gen}
\end{listing}

\begin{tcolorbox} [width=\linewidth, colback=yellow!30!white, top=1pt, bottom=1pt, left=2pt, right=2pt]
\textbf{Finding 1.2:}~\acp{LLM} have a high likelihood of producing invalid (\ie{} with syntactic errors) workflows with detailed prompts.
\end{tcolorbox}

\subsubsection{Final Evaluation}

\begin{figure}[ht]

    \begin{adjustbox}{right=8cm}
    \begin{tikzpicture}
    \begin{axis}[
        xbar,
        xmin=50,
        xmax=110,
        grid=major,
        major grid style={dashed},
        bar width=.2cm,
        width=8cm,
        height=8cm,
        legend style={at={(0.5,-0.1)},anchor=north,legend columns=-1},
        symbolic y coords={GPT-3.5 0-shot,CodeLlama 0-shot,StarChat 0-shot,GPT-3.5 1-shot,CodeLlama 1-shot,StarChat 1-shot,GPT-3.5F,CodeLlamaF,StarChatF},
        ytick=data,
        yticklabel style={text width=1.5cm,align=right},
        nodes near coords,
        nodes near coords align={horizontal},
        every node near coord/.append style={font=\small},
        ]
    \draw [purple!20, fill] (axis description cs:0, 0.66) rectangle (axis description cs:1,0.97); 
    \draw [orange!20, fill] (axis description cs:0, 0.35) rectangle (axis description cs:1,0.65); 
    \draw [yellow!20, fill] (axis description cs:0, 0.03) rectangle (axis description cs:1,0.34); 
    \addplot+ coordinates {(77.57,GPT-3.5 0-shot) (75.52,CodeLlama 0-shot) (67.81,StarChat 0-shot) (78.91,GPT-3.5 1-shot) (78.18,CodeLlama 1-shot) (74.67,StarChat 1-shot) (82.42,GPT-3.5F) (82.77,CodeLlamaF) (81.32,StarChatF)};
    \addplot+ coordinates {(80.68,GPT-3.5 0-shot) (92.71,CodeLlama 0-shot) (90.41,StarChat 0-shot) (90.34,GPT-3.5 1-shot) (93.31,CodeLlama 1-shot) (87.52,StarChat 1-shot) (95.38,GPT-3.5F) (97.71,CodeLlamaF) (90.60,StarChatF)};
    
    \legend{~\bleu{}(\%), ~\acck{}(\%)}
    \end{axis}
    \end{tikzpicture}
    \end{adjustbox}
    \caption{Final evaluation for workflow generation}
        \label{fig:wfsgen}
\end{figure}
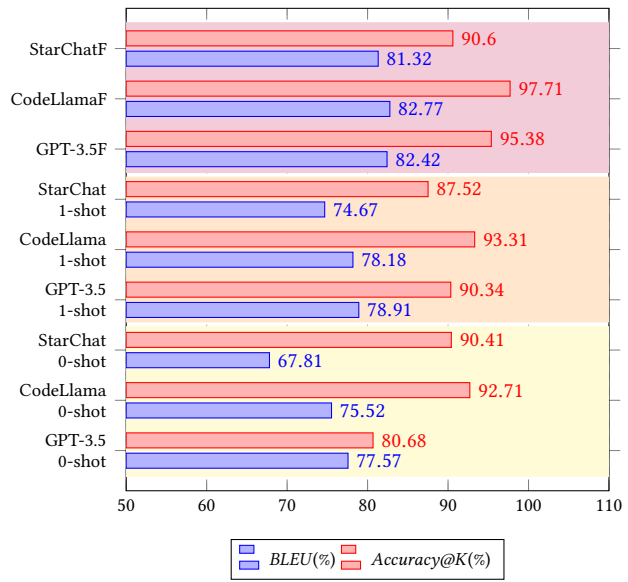

We selected the best configuration of each \ac{LLM} across different modes and performed our final evaluation.
The \fig{fig:wfsgen} shows the cumulative results across different modes.

\begin{tcolorbox} [width=\linewidth, colback=yellow!30!white, top=1pt, bottom=1pt, left=2pt, right=2pt]
\textbf{Finding 1.3:} For all~\acp{LLM}, the fine-tuned variant (\ie{} with F suffix) performs better than the corresponding off-the-shelf variant.
For all~\acp{LLM}, except for~\starchat{}, one-shot mode performs better than zero-shot.
\end{tcolorbox}

\noindent\emph{Effectiveness in generating expected workflows:} Higher~\bleu{} score indicates greater similarity between the generated and expected workflow.
For off-the-shelf variants,~\gpt{} achieves the best~\bleu{} score across all the modes. 
For fine-tuned variants,~\codellama{} has the best~\bleu{} scores.
Our analysis against the size of workflows (in~\apdx{apdx:sizeworkflowgeneration}) reveals that~\bleu{} scores do not vary with the size of workflows.
\begin{tcolorbox} [width=\linewidth, colback=yellow!30!white, top=1pt, bottom=1pt, left=2pt, right=2pt]
\textbf{Finding 1.4:} The ability of~\acp{LLM} to generate expected workflows does not vary much with the size of workflows.
\end{tcolorbox}

\noindent\emph{Ability to generate valid (\ie{} syntactically correct) workflows:}
Higher~\acck{} score indicates a greater chance of generating valid workflows.
For off-the-shelf and fine-tuned variants,~\codellama{} achieves the best~\acck{} score across all the modes.
Unlike~\bleu{} scores, the~\acck{} scores slowly decrease as workflow size becomes larger (details in~\apdx{apdx:sizeworkflowgeneration}).
This is expected as workflow size increases,~\acp{LLM} need to generate more tokens, increasing the likelihood of generating syntactic errors resulting in lower~\acck{} scores.

\begin{tcolorbox} [width=\linewidth, colback=green!30!white, top=1pt, bottom=1pt, left=2pt, right=2pt]
\textbf{Summary of RQ1:} Although~\gpt{} is highly likely to produce expected workflows. It might produce invalid or defective workflows.
On the other hand,~\codellama{} has a lesser likelihood of generating expected workflows but has a high probability of generating valid workflows.
\end{tcolorbox}

\noindent\emph{Effectiveness in generating semantically correct workflows:}
We want to assess~\acp{LLM} capability to generate semantically correct workflows.
However, automatically determining semantic correctness is impossible.
We decided to perform a manual validation by random sampling 30 valid workflows for each model and mode's best-performing~\bleu{} configurations.
In total, we manually checked 270 (30*9) workflows and computed the percentage of correctness for each workflow,~\ie{} percentage of generated steps that are semantically correct.
The~\tbl{tab:manual} shows the percentage of generated workflows with 100\% semantic correctness for each model. 
30 samples generated by \gpt{} across all the modes and~\codellama{} in 1-shot mode are all semantically correct workflows.
Other models also reach a high percentage.
The results indicate that generated workflows have a high semantic correctness. 

\begin{table}[h]
\caption{The percentage of workflows with 100\% semantic correctness.}
\vspace{-9pt}
\label{tab:manual}
%\scriptsize
\small
 
\begin{tabular}{c|cc|c}
\hline
\multirow{2}{*}{\textbf{Model}} & \multicolumn{2}{c|}{\textbf{off-the-shelf}}            & \multirow{2}{*}{\textbf{fine-tuned}}  \\ \cline{2-3}
                                & \multicolumn{1}{c|}{\textbf{0-shot}} & \textbf{1-shot} &                                     \\ \hline
\textbf{GPT-3.5}                & \multicolumn{1}{c|}{100\%}          & 100\%         & 100\%      \\ \hline
\textbf{CodeLlama}              & \multicolumn{1}{c|}{97.67\%}          & 100\%        & 86.67\%      \\ \hline
\textbf{StarChat}               & \multicolumn{1}{c|}{86.67\%}          & 93.33\%         &   96.66\% \\ \hline
\end{tabular}
\end{table}

\noindent\emph{How Secure are the Generated Workflows?}
Here, we want to evaluate how secure are the workflows generated by~\acp{LLM}.
Specifically, the number of syntactically valid workflows generated by~\acp{LLM} containing security issues.
We run~\argus{} on each of the workflows to detect security issues.
The~\tbl{tab:security_analysis} shows the results.
\starchat{} produced the most number of insecure workflows while~\gpt{} produced the least.
The~\lst{listing:generated_wf_with_vulns} shows an example of a workflow generated by~\gpt{} that has a code injection vulnerability. 
\begin{listing}[h]
\begin{minted}[breaklines,escapeinside=&&,fontsize=\scriptsize]{yaml}
name: Receive PR
on:
  pull_request:
    branches:
      - main
  workflow_dispatch:
jobs:
  test-pr:
    runs-on: ubuntu-latest
     ...
      - name: Set Outputs
        id: set-outputs
        run: &\textcolor{red}{\faBug}& echo ""::set-output name=is_valid::${{ steps.check-pr.outputs.VALID }}\n::set-output name=MSG::${{ steps.check-pr.outputs.MSG }}""
  save-pr-number:
    needs: test-pr
    runs-on: ubuntu-latest
    ...
\end{minted}
\caption{Example of a workflow generated by~\gpt{} that has a code injection vulnerability. The output~\textit{check-pr.outputs.MSG} is tainted (\ie{} controlled by non-repository owner). }
\label{listing:generated_wf_with_vulns}
\end{listing}

% Please add the following required packages to your document preamble:
% \usepackage{multirow}
\begin{table}[h]
%\caption{The number of syntactically valid workflows containing security issues.\machiry{Change this to number.} \xinyu{Fixed. format: a/b (c), a: the number of insecure workflows, b: the number of syntactically valid workflows, c: the number of syntactically valid workflows failing to tested on argus. Fine-tuned LLMs producing more insecure workflows may be caused by the fact that finetuning dataset contains many insecure workflows. Even though our objective of mixing fine-tuning dataset is to make LLMs understand multiple workflow-related tasks, it seems that it increases the likelyhood of generating insecure workflows.  }}
\caption{The number of syntactically valid workflows containing security issues.}
\vspace{-9pt}
\label{tab:security_analysis}
%\scriptsize
 \small
\begin{tabular}{c|cc|c|c}
\hline
\multirow{2}{*}{\textbf{Model}} & \multicolumn{2}{c|}{\textbf{off-the-shelf}}            & \multirow{2}{*}{\textbf{fine-tuned}} & \multirow{2}{*}{\textbf{Total}} \\ \cline{2-3}
                                & \multicolumn{1}{c|}{\textbf{0-shot}} & \textbf{1-shot} &    &                                  \\ \hline
\textbf{GPT-3.5}                & \multicolumn{1}{c|}{21}          & 10         & 66                             & 97\\ \hline
\textbf{CodeLlama}              & \multicolumn{1}{c|}{26}          & 35         & 213                             &  274\\ \hline
\textbf{StarChat}               & \multicolumn{1}{c|}{42}          & 51         & 252                              & 345\\ \hline
\end{tabular}

\end{table}
\begin{tcolorbox} [width=\linewidth, colback=yellow!30!white, top=1pt, bottom=1pt, left=2pt, right=2pt]
\textbf{Finding 1.5:}~\acp{LLM} can produce workflows with code injection vulnerabilities.
Developers should be careful while using~\ac{LLM} generated workflows.
\end{tcolorbox}

\subsection{RQ2: Defect Detection}
\label{subsec:rq2results}
As mentioned before, we are interested in~\acp{LLM} capability to detect two types of defects: syntactic errors and code injection vulnerabilities.
As mentioned in~\sect{subsec:githubworkflows}, detecting syntactic errors requires reasoning about the format of workflows. In other words, a well-formatted and syntactically valid~\code{yaml} can be an invalid workflow.
As mentioned in~\sect{subsubsec:evalmetrics}, we use~\fscore{} to measure detection capability and~\acck{} to measure detection accuracy (\ie{} line number).

\subsubsection{Syntactic Error Identification (T2)}
We evaluate this task using two prompts with varying details (\tbl{tab:methodology}).

\noindent\emph{Calibration:}
\tbl{tab:wfsbugfind_step1} shows the~\fscore{} and~\acck{} of different models and their variants across different modes.
Unlike workflow generation tasks, detailed prompts (P1 v/s P2) seem to have less effect on syntactic error detection.
{
\renewcommand{\arraystretch}{1.1} % Increase the row spacing
\setlength{\tabcolsep}{6pt} % Increase the column spacing
\begin{table}[h]
    \centering
    \caption{Effectiveness of syntactic error detection on~\ciset{}.}
    \vspace{-9pt}
  \resizebox{\columnwidth}{!}{% 
    \begin{tabular}{|c|*{16}{c|}}
    \hline
    %\multirow{2}{*}{\textbf{Model}} & \multirow{2}{*}{\textbf{t}} & \multicolumn{2}{c|}{\textbf{0-shot}} & \multicolumn{2}{c|}{\textbf{1-shot}} & \multicolumn{2}{c|}{\textbf{finetuned}} & & \multicolumn{2}{c|}{\textbf{0-shot}} & \multicolumn{2}{c|}{\textbf{1-shot}} & \multicolumn{2}{c|}{\textbf{finetuned}} \\
    %\cline{3-15}
    %& & P1 & P2 & P1 & P2 & P1 & P2 & & P1 & P2 & P1 & P2 & P1 & P2 \\ \hline
    \multirow{4}{*}{\textbf{Model}} & \multirow{4}{*}{\textbf{t}} & \multicolumn{6}{c|}{\textbf{F1-Score}}                                                                                                                                                     & \multirow{4}{*}{\textbf{}} & \multicolumn{6}{c|}{\textbf{Accuracy@K}}                                                                                                                                                   \\ \cline{3-8} \cline{10-15} 
                                &                             & \multicolumn{4}{c|}{\textbf{off-the-shelf}}                                                                                               & \multicolumn{2}{c|}{\multirow{2}{*}{\textbf{fine-tuned}}}       &                            & \multicolumn{4}{c|}{\textbf{off-the-shelf}}                                                                                               & \multicolumn{2}{c|}{\multirow{2}{*}{\textbf{fine-tuned}}}       \\ \cline{3-6} \cline{10-13} 
                                &                             & \multicolumn{2}{c|}{\textbf{0-shot}}                                & \multicolumn{2}{c|}{\textbf{1-shot}}                                & \multicolumn{2}{c|}{\textbf{}}           &                            & \multicolumn{2}{c|}{\textbf{0-shot}}                                & \multicolumn{2}{c|}{\textbf{1-shot}}                                & \multicolumn{2}{c|}{\textbf{}}           \\ \cline{3-8} \cline{10-15} 
                                &                             & \multicolumn{1}{c|}{\textbf{P1}} & \multicolumn{1}{c|}{\textbf{P2}} & \multicolumn{1}{c|}{\textbf{P1}} & \multicolumn{1}{c|}{\textbf{P2}} & \multicolumn{1}{c|}{\textbf{P1}} & \textbf{P2} &                            & \multicolumn{1}{c|}{\textbf{P1}} & \multicolumn{1}{c|}{\textbf{P2}} & \multicolumn{1}{c|}{\textbf{P1}} & \multicolumn{1}{c|}{\textbf{P2}} & \multicolumn{1}{c|}{\textbf{P1}} & \textbf{P2} \\ \hline

    \multirow{5}{*}{\textbf{\gpt{}}}
    & 0.1 & \shadecell{61.40} & \shadeunderlinecell{72.25} & \shadecell{3.170} & \shadecell{2.560} & \shadecell{87.76} & \shadecell{90.45} & & \shadecell{27.00} & \shadecell{39.00} & \shadecell{3.000} & \shadecell{1.000} & \shadecell{81.00} & \shadecell{85.00} \\
    & 0.3 & \shadecell{56.62} & \shadecell{69.43} & \shadecell{3.230} & \shadecell{1.270} & \shadecell{88.21} & \shadecell{90.45} & & \shadecell{37.00} & \shadecell{41.00} & \shadecell{4.000} & \shadecell{3.000} & \shadecell{83.00} & \shadecell{86.00} \\
    & 0.5 & \shadecell{58.56} & \shadecell{68.69} & \shadecell{4.260} & \shadecell{3.730} & \shadecell{87.18} & \shadecell{90.45} & & \shadecell{39.00} & \shadecell{42.00} & \shadecell{5.000} & \shadecell{5.000} & \shadecell{83.00} & \shadecell{86.00} \\
    & 0.7 & \shadecell{62.78} & \shadecell{70.41} & \shadeunderlinecell{4.320} & \shadecell{1.270} & \shadecell{88.21} & \shadeunderlinecell{91.46} & & \shadeunderlinecell{46.00} & \shadecell{44.00} & \shadecell{7.000} & \shadeunderlinecell{8.000} & \shadecell{86.00} & \shadecell{89.00} \\
    & 0.9 & \shadecell{54.13} & \shadecell{65.98} & \shadecell{3.240} & \shadecell{3.800} & \shadecell{86.87} & \shadecell{90.55} & & \shadecell{42.00} & \shadecell{43.00} & \shadecell{6.000} & \shadecell{6.000} & \shadecell{86.00} & \shadeunderlinecell{90.00} \\ \hline
    \multirow{5}{*}{\textbf{\codellama{}}}
    & 0.1 & \shadecell{17.09} & \shadecell{31.88} & \shadeunderlinecell{32.00} & \shadecell{7.270} & \shadecell{71.97} & \shadecell{80.37} & & \shadecell{1.000} & \shadecell{6.000} & \shadecell{1.000} & \shadecell{0.000} & \shadecell{48.00} & \shadecell{51.00} \\
    & 0.3 & \shadecell{33.09} & \shadecell{37.42} & \shadecell{25.21} & \shadecell{3.740} & \shadecell{72.80} & \shadeunderlinecell{80.75} & & \shadecell{1.000} & \shadecell{8.000} & \shadecell{2.000} & \shadecell{2.000} & \shadecell{51.00} & \shadecell{55.00} \\
    & 0.5 & \shadecell{45.16} & \shadecell{44.30} & \shadecell{25.21} & \shadecell{9.090} & \shadecell{70.39} & \shadecell{79.64} & & \shadecell{4.000} & \shadeunderlinecell{11.00} & \shadeunderlinecell{8.000} & \shadecell{5.000} & \shadecell{50.00} & \shadeunderlinecell{59.00} \\
    & 0.7 & \shadeunderlinecell{46.05} & \shadecell{40.99} & \shadecell{25.81} & \shadecell{3.850} & \shadecell{70.18} & \shadecell{79.82} & & \shadecell{3.000} & \shadeunderlinecell{11.00} & \shadecell{6.000} & \shadecell{3.000} & \shadecell{54.00} & \shadecell{57.00} \\
    & 0.9 & \shadecell{40.25} & \shadecell{41.51} & \shadecell{26.45} & \shadecell{7.340} & \shadecell{70.94} & \shadecell{78.57} & & \shadecell{2.000} & \shadecell{10.00} & \shadeunderlinecell{8.000} & \shadecell{5.000} & \shadeunderlinecell{59.00} & \shadecell{56.00} \\ \hline
    \multirow{5}{*}{\textbf{\starchat{}}}
    & 0.1 & \shadecell{67.34} & \shadecell{66.67} & \shadeunderlinecell{100.0} & \shadeunderlinecell{100.0} & \shadecell{64.20} & \shadeunderlinecell{84.47} & & \shadecell{6.000} & \shadecell{12.00} & \shadecell{12.00} & \shadecell{10.00} & \shadecell{49.00} & \shadecell{62.00} \\
    & 0.3 & \shadecell{67.34} & \shadecell{68.03} & \shadeunderlinecell{100.0} & \shadeunderlinecell{100.0} & \shadecell{62.65} & \shadecell{84.16} & & \shadecell{10.00} & \shadecell{13.00} & \shadecell{17.00} & \shadecell{14.00} & \shadecell{52.00} & \shadecell{63.00} \\
    & 0.5 & \shadecell{65.68} & \shadeunderlinecell{69.82} & \shadecell{98.49} & \shadecell{99.50} & \shadecell{68.26} & \shadecell{82.76} & & \shadeunderlinecell{15.00} & \shadecell{14.00} & \shadecell{13.00} & \shadecell{20.00} & \shadecell{55.00} & \shadecell{65.00} \\
    & 0.7 & \shadecell{51.16} & \shadecell{60.68} & \shadecell{96.04} & \shadecell{97.98} & \shadecell{65.90} & \shadecell{83.58} & & \shadecell{10.00} & \shadecell{14.00} & \shadecell{14.00} & \shadeunderlinecell{29.00} & \shadecell{54.00} & \shadeunderlinecell{69.00} \\
    & 0.9 & \shadecell{47.13} & \shadecell{57.45} & \shadecell{89.11} & \shadecell{91.98} & \shadecell{64.80} & \shadecell{80.98} & & \shadecell{7.000} & \shadecell{14.00} & \shadecell{21.00} & \shadecell{14.00} & \shadecell{51.00} & \shadecell{65.00} \\ \hline
    \end{tabular}
     }
    \label{tab:wfsbugfind_step1}
\end{table}
}

\noindent\emph{\fscore{}:}  
In 0-shot mode, \gpt{} performs the best in detecting syntactic errors with the highest~\fscore{} of 72.25\%.
The performance of~\gpt{} and~\codellama{} dropped in 1-shot mode --- contrary to previous works~\cite{baiexploring, 10.1145/3411763.3451760} which show that 1-shot mode provides better performance than 0-shot mode.
As expected, in~\gpt{} and~\codellama{}, the fine-tuned variants performed better than off-the-shelf variants.
The case is different for~\starchat{}, where the 1-shot mode of the off-the-shelf variant performed the best, even better than the fine-tuned variant.

\noindent\emph{\acck{}:} 
The detection accuracy of off-the-shelf variants follows the same trend as the detection capability. In other words, ~\gpt{} performs the best in 0-shot mode, and 1-shot mode hurts the performance of ~\gpt{} and~\codellama{} but improves that of~\starchat{}. As expected, fine-tuned variants perform better than off-the-shelf variants.

\noindent\emph{Final Evaluation:}
The~\fig{fig:wfsbugfind} (in Appendix) shows the evaluation of the best-performing configuration on the final large dataset.
Overall,~\starchat{} 1-shot mode is the best at detecting syntactic errors as indicated by the highest~\fscore{},~\ie{} 100\%.
However, fine-tuned~\gpt{} has the highest accuracy.
In other words,~\starchat{} is good at detecting whether a workflow has a syntactic error or not. But, fine-tuned~\gpt{} is good at detecting where (\ie{} line number) the syntactic error is.
\lst{listing:syntaxerror_example} (in Appendix) shows an example where~\starchat{} correctly identified a syntactic error but~\gpt{} failed.

\begin{tcolorbox} [width=\linewidth, colback=yellow!30!white, top=1pt, bottom=1pt, left=2pt, right=2pt]
\textbf{Finding 2.1:} Contrary to the observations for other applications, for~\gpt{} and~\codellama{}, the 1-shot mode is less effective than 0-shot in identifying syntactic errors in workflows.
\starchat{} is best at detecting syntactic errors but~\gpt{} can accurately identify the location of syntactic error.
\end{tcolorbox}

\subsubsection{Code Injection Vulnerability Detection (T4)}
As shown in~\tbl{tab:methodology}, we use three prompts to evaluate this task.

\noindent\emph{Calibration:} The~\tbl{tab:wfsvulfind_step1_f1score} shows the~\fscore{} of code injection vulnerability detection of different models and their variants across different modes.
The fine-tuned variants perform best for all~\acp{LLM} and a given prompt.
For off-the-shelf variants of~\codellama{} and~\starchat{}, simpler prompts (\ie{} P1 and P2) provide the best~\fscore{}.
However, for~\gpt{}, the detailed prompt (\ie{} P3) provides the best~\fscore{}.
For the off-the-shelf variants of~\codellama{} and~\starchat{}, smaller temperature values (\ie{} low non-determinism) provide the best~\fscore{}. In contrast, higher temperature (\ie{} higher non-determinism) works well for~\gpt{}.
As shown in~\tbl{tab:wfsvulfind_step1_acc} (in Appendix) the~\acck{} follows the expected trend,~\ie{} detailed prompt gives better results, and the fine-tuned variant performs better than the corresponding off-the-shelf variant.
Similar to~\fscore{}, simpler prompts (\ie{} P1 and P2) work well for~\codellama{} and~\starchat{}. Whereas detailed prompt (\ie{} P3) works well for~\gpt{}.

{
\renewcommand{\arraystretch}{1.1} % Increase the row spacing
\setlength{\tabcolsep}{6pt} % Increase the column spacing
\begin{table}[h]
    \centering
    \caption{F1 scores of code injection vulnerability detection on~\ciset{}.}
    \vspace{-9pt}
\resizebox{\columnwidth}{!}{% 
    \begin{tabular}{|c|*{11}{c|}}
    \hline
    \multirow{3}{*}{\textbf{Model}} & \multirow{3}{*}{\textbf{t}} & \multicolumn{6}{c|}{\textbf{off-the-shelf}} & \multicolumn{3}{c|}{\multirow{2}{*}{\textbf{fine-tuned}}} \\
    \cline{3-8}
    & & \multicolumn{3}{c|}{\textbf{0-shot}} & \multicolumn{3}{c|}{\textbf{1-shot}} & \multicolumn{3}{c|}{\textbf{}}\\
    \cline{3-11}
    & & P1 & P2 & P3 & P1 & P2 & P3 & P1 & P2 & P3 \\ \hline
    \multirow{5}{*}{\textbf{\gpt{}}}
    & 0.1 & \shadecell{7.140} & \shadecell{17.98} & \shadecell{80.00} & \shadecell{1.770} & \shadecell{0.000} & \shadecell{24.14} & \shadecell{93.49} & \shadecell{92.86} & \shadeunderlinecell{99.38} \\
    & 0.3 & \shadecell{7.140} & \shadecell{20.00} & \shadecell{78.57} & \shadecell{0.000} & \shadecell{0.000} & \shadecell{25.42} & \shadecell{92.94} & \shadecell{92.40} & \shadeunderlinecell{99.38} \\
    & 0.5 & \shadecell{9.410} & \shadecell{21.98} & \shadecell{79.52} & \shadecell{0.000} & \shadecell{0.000} & \shadecell{24.78} & \shadecell{92.94} & \shadecell{92.31} & \shadeunderlinecell{99.38} \\
    & 0.7 & \shadecell{2.440} & \shadecell{17.78} & \shadeunderlinecell{81.93} & \shadecell{0.000} & \shadecell{6.520} & \shadecell{25.86} & \shadecell{92.94} & \shadecell{92.31} & \shadeunderlinecell{99.38} \\
    & 0.9 & \shadecell{11.63} & \shadecell{17.78} & \shadecell{79.04} & \shadecell{5.130} & \shadecell{0.000} & \shadeunderlinecell{27.12} & \shadecell{93.49} & \shadecell{93.49} & \shadeunderlinecell{99.38} \\ \hline
    \multirow{5}{*}{\textbf{\codellama{}}}
    & 0.1 & \shadecell{66.11} & \shadecell{68.67} & \shadecell{66.67} & \shadeunderlinecell{87.86} & \shadecell{74.88} & \shadecell{66.67} & \shadecell{89.53} & \shadeunderlinecell{89.66} & \shadecell{88.34} \\
    & 0.3 & \shadecell{66.10} & \shadecell{67.54} & \shadecell{66.67} & \shadecell{85.54} & \shadecell{73.89} & \shadecell{66.67} & \shadecell{89.02} & \shadecell{89.02} & \shadecell{87.80} \\
    & 0.5 & \shadecell{63.72} & \shadeunderlinecell{68.81} & \shadecell{66.67} & \shadecell{77.50} & \shadecell{70.59} & \shadecell{67.80} & \shadecell{87.78} & \shadecell{89.02} & \shadecell{88.20} \\
    & 0.7 & \shadecell{60.66} & \shadecell{59.11} & \shadecell{67.56} & \shadecell{74.17} & \shadecell{67.05} & \shadecell{68.09} & \shadecell{84.39} & \shadecell{87.86} & \shadecell{87.80} \\
    & 0.9 & \shadecell{63.64} & \shadecell{67.01} & \shadecell{67.59} & \shadecell{69.74} & \shadecell{56.18} & \shadecell{70.27} & \shadecell{86.86} & \shadecell{88.24} & \shadecell{84.66} \\ \hline
    \multirow{5}{*}{\textbf{\starchat{}}}
    & 0.1 & \shadecell{66.67} & \shadecell{66.67} & \shadecell{66.67} & \shadecell{72.07} & \shadecell{87.43} & \shadecell{72.90} & \shadecell{94.41} & \shadecell{93.83} & \shadecell{96.86} \\
    & 0.3 & \shadecell{66.67} & \shadecell{66.11} & \shadecell{66.67} & \shadecell{72.07} & \shadeunderlinecell{89.89} & \shadecell{72.48} & \shadecell{94.41} & \shadecell{95.00} & \shadecell{96.20} \\
    & 0.5 & \shadecell{66.67} & \shadeunderlinecell{67.24} & \shadecell{65.25} & \shadecell{76.56} & \shadecell{84.66} & \shadecell{73.49} & \shadecell{93.08} & \shadecell{93.17} & \shadecell{95.54} \\
    & 0.7 & \shadecell{66.09} & \shadecell{64.55} & \shadecell{65.50} & \shadecell{77.61} & \shadecell{88.89} & \shadecell{73.93} & \shadecell{93.67} & \shadecell{93.83} & \shadeunderlinecell{97.50} \\
    & 0.9 & \shadecell{59.81} & \shadecell{56.84} & \shadecell{60.00} & \shadecell{74.40} & \shadecell{84.44} & \shadecell{75.12} & \shadecell{91.82} & \shadecell{92.59} & \shadeunderlinecell{97.50} \\ \hline
    \end{tabular}
        }

    \label{tab:wfsvulfind_step1_f1score}
\end{table}
}
\begin{figure}[ht]
    \begin{adjustbox}{right=8cm}
    \begin{tikzpicture}
    \begin{axis}[
        xbar,
        xmin=0,
        xmax=115,
        grid=major,
        major grid style={dashed},
        bar width=.2cm,
        width=8cm,
        height=8cm,
        legend style={at={(0.5,-0.1)},anchor=north,legend columns=-1},
        symbolic y coords={GPT-3.5 0-shot,CodeLlama 0-shot,StarChat 0-shot,GPT-3.5 1-shot,CodeLlama 1-shot,StarChat 1-shot,GPT-3.5F,CodeLlamaF,StarChatF},
        ytick=data,
        yticklabel style={text width=1.5cm,align=right},
        nodes near coords,
        nodes near coords align={horizontal},
        every node near coord/.append style={font=\small},
        ]
    \draw [purple!20, fill] (axis description cs:0, 0.66) rectangle (axis description cs:1,0.97); 
    \draw [orange!20, fill] (axis description cs:0, 0.35) rectangle (axis description cs:1,0.65); 
    \draw [yellow!20, fill] (axis description cs:0, 0.03) rectangle (axis description cs:1,0.34); 
    \addplot+ coordinates {(76.15,GPT-3.5 0-shot) (65.54,CodeLlama 0-shot) (66.47,StarChat 0-shot) (29.27,GPT-3.5 1-shot) (86.52,CodeLlama 1-shot) (88.75,StarChat 1-shot) (98.82,GPT-3.5F) (86.93,CodeLlamaF) (96.47,StarChatF)};
    
    \addplot+ coordinates {(8.00,GPT-3.5 0-shot) (2.87,CodeLlama 0-shot) (0.59,StarChat 0-shot) (4.63,GPT-3.5 1-shot) (0.0,CodeLlama 1-shot) (23.00,StarChat 1-shot) (82.52,GPT-3.5F) (72.15,CodeLlamaF) (78.30,StarChatF)};
    
    \legend{~\fscore{}(\%), ~\acck{}(\%)}
    \end{axis}
    \end{tikzpicture}
    \end{adjustbox}
    \caption{Final evaluation for code injection vulnerability detection.}
    \label{fig:vulfind}
\end{figure}
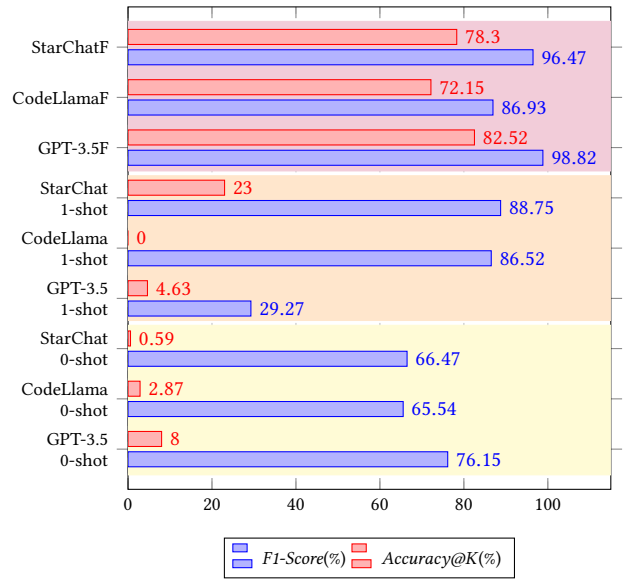

\noindent\emph{Final Evaluation:} The~\fig{fig:vulfind} shows the evaluation of the best-performing configuration on the final large dataset.
Overall, fine-tuned variants perform better, demonstrating the importance of fine-tuning in detecting code injection vulnerabilities.
The fine-tuned variant of~\gpt{} (\ie{}GPT-3.5F) performs the best.
\emph{Interestingly, off-the-shelf~\gpt{} performs the worst in 1-shot mode.}

\lst{lst:gptf_vuln_dection} (in Appendix) shows an example where~\gpt{} correctly identified a code injection vulnerability missed by other~\acp{LLM}.
We discuss the detection effectiveness against the size of workflows in~\apdx{apdx:sizedefectdetection}.
The detection effectiveness (both syntactic errors and code injection vulnerabilities) of most \acp{LLM} does not vary with the size of workflows.

\begin{tcolorbox} [width=\linewidth, colback=green!30!white, top=1pt, bottom=1pt, left=2pt, right=2pt]
\textbf{Summary of RQ2:} Across the tested~\acp{LLM}, there is a significant difference in the effectiveness of syntactic error detection and code injection vulnerability detection.
Off-the-shelf~\starchat{} in 1-shot mode is best at detecting syntactic errors, whereas fine-tuned~\gpt{} is best at detecting code injection vulnerabilities.
\end{tcolorbox}

\subsection{RQ3: Defect Repair}
\label{subsec:rq3results}
Similar to defect detection, we focus on repairing two kinds of defects: syntactic errors and code injection vulnerabilities.
We use~\acck{} to assess the effectiveness of defect repair.
As described in~\sect{subsec:instrfinetuning}, we do not include repair examples in our fine-tuning dataset.
Hence defect repairs (T4 and T5) can be considered as unseen (but related) tasks for~\acp{LLM}.

\subsubsection{Syntactic Error Repair (T4)}
We evaluate this task using three prompts (P1, P2, P3) with increasing detail (\tbl{tab:methodology}).

\noindent{\emph{Calibration:}} The~\tbl{tab:wfsbugfix_step1_static_check_accuracy} shows the~\acck{} of different~\acp{LLM} on our calibration dataset (\ciset{}).
Across all prompts, higher temperatures yield better results.
This is expected as higher temperature value allows~\acp{LLM} to be more creative, consequently increasing the likelihood of generating repaired workflow.
The~\lst{code:repair_eg} shows an example of a syntactically invalid workflow due to the use of a invalid step name (\textcolor{syntxbugcolor}{\faBug}).
In this instance, setting a higher temperature value successfully corrected the syntactic error, whereas a lower temperature setting failed to do so.
\begin{listing}[t]
\begin{minted}[breaklines,escapeinside=&&,fontsize=\tiny]{yaml}
name: Build and deploy Python app to Azure Web App - cloudnew
on:
  push:
    branches:
      - main
  workflow_dispatch:
jobs:
  build:
    runs-on: ubuntu-latest
    steps:
       ...
  deploy:
    runs-on: ubuntu-latest
    needs: build
    environment:
      name: 'production'
      & \textcolor{syntxbugcolor}{\faBug} & # url: ${{ steps.deploy-to-webapp.outputs.webapp-url }} 
    steps:
      - name: Download artifact from build job
        uses: actions/download-artifact@v2
        with:
          name: python-app
          path: .
      - name: 'Deploy to Azure Web App'
        uses: azure/webapps-deploy@v2
        with:
          app-name: 'cloudnew'
          slot-name: 'production'
          publish-profile: ${{ secrets.AzureAppService_PublishProfile_3692f3e0dfb044c19960d501da0e7555 }}
\end{minted}
\caption{A syntactically invalid workflow was successfully fixed by~\gpt{} at temperature 0.9, but~\gpt{} with the same setting at temperature 0.1 failed to repair it.}
\label{code:repair_eg}
\end{listing}

%     - uses: actions/checkout@v2
%    - name: Set up Python version
%      uses: actions/setup-python@v1
%      with:
%        python-version: '3.7'
%    - name: Create and start virtual environment
%      run: |
%        python -m venv venv
%        source venv/bin/activate
%    - name: Install dependencies
%      run: pip install -r requirements.txt
%    - name: Upload artifact for deployment jobs
%      uses: actions/upload-artifact@v2
%      with:
%        name: python-app
%        path: |
%          .
%          !venv/

Also, detailed prompts provide better results, as indicated by the increasing trend across P1 to P3.
For simpler prompts,~\ie{} P1 and P2, fine-tuned variant of~\gpt{} perform better on syntactic error repair tasks (unseen tasks) than the off-the-shelf variant.
However, the case is different with~\codellama{} and~\starchat{}, where the fine-tuned variant performed poorly.
These results demonstrate that fine-tuning~\gpt{} on certain tasks helps in improving its effectiveness on other unseen but related tasks.
However, this is not the case with other~\acp{LLM}, where fine-tuned variants can perform poorly on unseen (but related) tasks.
Intuitively, this makes sense as~\gpt{} is trained on diverse datasets and has higher generalization capability. Whereas specialized~\acp{LLM} (\ie{}~\codellama{} and~\starchat{}) have less generalization capability.
Our observations are in line with prior work~\cite{wei2022finetuned}, which showed that fine-tuned large models (\eg{}~\gpt{}) generalize to unseen (but related) tasks. In contrast, smaller models (\eg{}~\codellama{} and~\starchat{}) suffer as all model capacity is used for tasks used in fine-tuning.

\noindent{\emph{Final Evaluation:}}
The~\fig{fig:wfsbugfix_step2} (in Appendix) shows $Accuracy@k$ of syntactic error fixing on the large dataset. We did not include the results for fine-tuned variants of~\codellama{} and~\starchat{} as they are extremely poor (\ie{} $< 40\%$).
\gpt{} in 1-shot mode performs the best across all~\acp{LLM} and their variants.

{
\renewcommand{\arraystretch}{1.1} % Increase the row spacing
\setlength{\tabcolsep}{6pt} % Increase the column spacing
\begin{table}[t]
    \centering
    \caption{Accuracy@K of syntax error fixing on \ciset{}.}
    \vspace{-9pt}
\resizebox{\columnwidth}{!}{% 
    \begin{tabular}{|c|*{14}{c|}}
    \hline
    \multirow{3}{*}{\textbf{Model}} & \multirow{3}{*}{\textbf{t}} & \multicolumn{6}{c|}{\textbf{off-the-shelf}} & \multicolumn{6}{c|}{\textbf{fine-tuned}} \\
    \cline{3-14}
    & & \multicolumn{3}{c|}{\textbf{0-shot}} & \multicolumn{3}{c|}{\textbf{1-shot}} & \multicolumn{3}{c|}{\textbf{0-shot}} & \multicolumn{3}{c|}{\textbf{1-shot}}\\
    \cline{3-14}
    & & P1 & P2 & P3 & P1 & P2 & P3 & P1 & P2 & P3 & P1 & P2 & P3 \\ \hline
    \multirow{5}{*}{\textbf{\gpt{}}}
    & 0.1 & \shadecell{46.50} & \shadecell{50.00} & \shadecell{80.50} & \shadecell{51.27} & \shadecell{56.85} & \shadecell{82.23} & \shadecell{65.00} & \shadecell{65.50} & \shadecell{82.50} & \shadecell{70.53} & \shadecell{59.26} & \shadecell{80.65} \\
    & 0.3 & \shadecell{47.50} & \shadecell{52.50} & \shadecell{82.50} & \shadecell{52.79} & \shadecell{57.87} & \shadecell{84.77} & \shadecell{67.00} & \shadecell{69.50} & \shadecell{87.00} & \shadecell{72.63} & \shadecell{65.08} & \shadecell{85.48} \\
    & 0.5 & \shadecell{48.50} & \shadecell{52.50} & \shadecell{86.00} & \shadecell{54.31} & \shadecell{59.90} & \shadecell{90.86} & \shadecell{67.50} & \shadecell{71.00} & \shadecell{89.00} & \shadecell{74.21} & \shadecell{68.78} & \shadecell{86.56} \\
    & 0.7 & \shadecell{50.00} & \shadecell{57.00} & \shadecell{88.50} & \shadecell{55.33} & \shadecell{61.42} & \shadecell{93.40} & \shadecell{68.50} & \shadecell{75.50} & \shadecell{90.00} & \shadecell{71.21} & \shadecell{71.43} & \shadecell{87.63} \\
    & 0.9 & \shadecell{53.00} & \shadecell{58.00} & \shadeunderlinecell{91.50} & \shadecell{56.85} & \shadecell{66.50} & \shadeunderlinecell{93.91} & \shadecell{73.00} & \shadecell{75.50} & \shadeunderlinecell{91.00}  & \shadecell{76.32} & \shadecell{73.54} & \shadeunderlinecell{88.17} \\ \hline
    \multirow{5}{*}{\textbf{\codellama{}}}
    & 0.1 & \shadecell{4.040} & \shadecell{17.17} & \shadecell{62.12} & \shadecell{9.000} & \shadecell{52.50} & \shadecell{71.50} & \shadecell{0.000} & \shadecell{0.510} & \shadecell{0.510}  & \shadecell{5.000} & \shadecell{7.000} & \shadecell{7.500} \\
    & 0.3 & \shadecell{3.540} & \shadecell{36.87} & \shadecell{87.88} & \shadecell{11.50} & \shadecell{35.00} & \shadecell{88.50} & \shadecell{0.000} & \shadecell{0.510} & \shadecell{0.000}  & \shadecell{11.50} & \shadecell{10.50} & \shadecell{12.00} \\
    & 0.5 & \shadecell{6.060} & \shadecell{35.86} & \shadecell{87.37} & \shadecell{14.00} & \shadecell{33.00} & \shadecell{86.00} & \shadecell{0.000} & \shadecell{0.000} & \shadecell{1.520}  & \shadecell{17.50} & \shadecell{15.00} & \shadecell{17.00} \\
    & 0.7 & \shadecell{8.590} & \shadecell{40.91} & \shadeunderlinecell{89.39} & \shadecell{18.00} & \shadecell{33.00} & \shadecell{86.50} & \shadecell{2.530} & \shadecell{1.520} & \shadecell{2.530}  & \shadecell{27.50} & \shadecell{23.00} & \shadecell{25.50} \\
    & 0.9 & \shadecell{15.15} & \shadecell{36.36} & \shadecell{86.36} & \shadecell{19.00} & \shadecell{37.00} & \shadeunderlinecell{92.50} & \shadecell{3.030} & \shadecell{3.030} & \shadeunderlinecell{3.540}  & \shadeunderlinecell{35.50} & \shadecell{25.50} & \shadecell{30.00} \\ \hline
    \multirow{5}{*}{\textbf{\starchat{}}}
    & 0.1 & \shadecell{44.44} & \shadecell{49.49} & \shadecell{60.10} & \shadecell{42.00} & \shadecell{43.50} & \shadecell{45.00} & \shadecell{0.000} & \shadecell{0.000} & \shadecell{1.010}  & \shadecell{0.000} & \shadecell{0.500} & \shadecell{0.000} \\
    & 0.3 & \shadecell{44.44} & \shadecell{51.01} & \shadecell{62.12} & \shadecell{43.50} & \shadecell{44.00} & \shadecell{47.00} & \shadecell{0.000} & \shadecell{0.510} & \shadecell{1.010}  & \shadecell{0.000} & \shadecell{2.500} & \shadecell{0.000} \\
    & 0.5 & \shadecell{44.95} & \shadecell{52.02} & \shadecell{64.14} & \shadecell{43.50} & \shadecell{44.50} & \shadecell{48.50} & \shadecell{0.000} & \shadecell{0.510} & \shadecell{1.010}  & \shadecell{0.000} & \shadecell{5.000} & \shadecell{0.000} \\
    & 0.7 & \shadecell{44.44} & \shadecell{53.54} & \shadeunderlinecell{64.65} & \shadecell{32.50} & \shadecell{29.50} & \shadecell{54.00} & \shadecell{0.000} & \shadecell{0.510} & \shadecell{1.520}  & \shadecell{0.500} & \shadecell{8.000} & \shadecell{0.000} \\
    & 0.9 & \shadecell{42.42} & \shadecell{52.02} & \shadeunderlinecell{64.65} & \shadecell{42.00} & \shadecell{36.50} & \shadeunderlinecell{55.00} & \shadecell{1.520} & \shadecell{0.000} & \shadeunderlinecell{2.530}  & \shadecell{1.000} & \shadeunderlinecell{9.000} & \shadecell{0.500} \\ \hline
    \end{tabular}
        }
    
    \label{tab:wfsbugfix_step1_static_check_accuracy}
\end{table}
}

\subsubsection{Code Injection Vulnerability Repair (T5)}
The effectiveness of repairing code injection vulnerabilities also follows a similar pattern as repairing syntactic errors.
Also, all~\acp{LLM} and their fine-tuned variants suffer in repairing code injection vulnerabilities.
More details are provided in~\apdx{apdx:codeinjecrepair}.

\begin{tcolorbox} [width=\linewidth, colback=green!30!white, top=1pt, bottom=1pt, left=2pt, right=2pt]
\textbf{Summary of RQ3:}~\acp{LLM} perform well (at higher temperatures) in repairing syntactic errors but suffers at repairing code injection vulnerabilities.
Fine-tuning~\codellama{} and~\starchat{} hurts their performance on unseen (but related) workflow tasks.
\end{tcolorbox}

\section{Threats to Validity} 
\label{sec:discussion}
We identified the following potential (generalizability) threats to the validity of our study. 
\begin{itemize}[leftmargin=*]
\item\textbf{Generalizability to Other Tasks:} We investigated three categories of tasks. However, there could be other related tasks (\eg{} refactoring) on which the effectiveness of~\acp{LLM} might differ. We tried to handle this in RQ3 (\sect{subsec:rq3results}), where all the tasks are unseen but related.
\item\textbf{Generalizability to Other~\acp{LLM}:}
We have investigated three~\acp{LLM}, and the observations may not generalize to other~\acp{LLM} that are architected differently.
Our datasets and experimentation scripts will enable easy evaluation of any given~\ac{LLM} and compare against our results.
\item\textbf{Generalizability to Other CI platforms:}
We anticipate that our observations will generalize to other CI platforms as well because most of the CI platforms follow the same syntax (\ie{}~\code{YAML}) and have a similar structure~\cite{characterizinggithub}.
\end{itemize}

\section{Conclusion}
\label{sec:conlusion}
We perform the first large-scale study to investigate the effectiveness of three state-of-the-art~\acp{LLM} and their fine-tuned variants on five tasks related to~\gw{}.
We curated a set of $\sim$400K workflows with various prompts with varying details across different tasks.
Our study revealed various interesting findings and open problems in using~\acp{LLM} for workflows.
For instance,~\acp{LLM} suffer at generating large and valid workflows.
\acp{LLM} are not effective at repairing code injection vulnerabilities.

% References need to start on a new page.
% Rules: https://www.ares-conference.eu/submission/
\pagebreak
%
% ---- Bibliography ----
\bibliographystyle{ACM-Reference-Format}
\bibliography{bibliography}

\pagebreak
\appendix
\section{Appendices}

\subsection{Workflow Complexity Metrics}
\label{apdx:workflowcomplex}
Based on the structure (\sect{subsec:githubworkflows}) of~\gw{}, we used six complexity metrics to gauge the complexity of a workflow,~\ie{} the number of triggers, the number of actions, the number of reusable workflows, the number of jobs, the number of steps and cyclomatic complexity. We bucketed each complexity metric and removed workflows containing outliers. \tbl{tab:complexitymetrics} presents the data distribution of all the remaining~\gw{} for each complexity metric. Each complexity metric of a workflow falls into a bucket, and we used the combination of complexity metrics as classification criteria to divide all workflows into different groups. We found that 133 combinations have over 100 workflows. Thus, we regard workflows belonging to these 133 combinations as representative workflows.
% Please add the following required packages to your document preamble:
% \usepackage{multirow}
\begin{table}[h]
\centering
\tiny
\caption{Data bucketing for complexity metrics (No. =  the number of)}
\label{tab:complexitymetrics}
\resizebox{\columnwidth}{!}{% 
\begin{tabular}{|c|c|c|c|}
\hline
\textbf{Complexity Metric}                                 & \textbf{Data Bucketing} & \textbf{No. Workflows} & \textbf{Proportion} \\ \hline
\multirow{3}{*}{\textbf{No. Triggers}}                     & 1                       & 157,470                              & 0.5410              \\ \cline{2-4} 
                                                           & 2                       & 92,960                               & 0.3194              \\ \cline{2-4} 
                                                           & 3-6                     & 40,646                               & 0.1396              \\ \hline
\multirow{5}{*}{\textbf{No. Actions}}                      & 0-1                     & 58,357                               & 0.2005              \\ \cline{2-4} 
                                                           & 2                       & 88,481                               & 0.3040              \\ \cline{2-4} 
                                                           & 3                       & 71,860                               & 0.2469              \\ \cline{2-4} 
                                                           & 4                       & 46,871                               & 0.1610              \\ \cline{2-4} 
                                                           & 5-10                    & 25,507                               & 0.0876              \\ \hline
\multirow{2}{*}{\textbf{No. Reusable Workflows}}           & 0                       & 277,513                              & 0.9534              \\ \cline{2-4} 
                                                           & 1-4                     & 13,563                               & 0.0466              \\ \hline
\multirow{3}{*}{\textbf{No. Jobs}}                         & 1                       & 256,486                              & 0.8812              \\ \cline{2-4} 
                                                           & 2-3                     & 31,579                               & 0.1085              \\ \cline{2-4} 
                                                           & 4-8                     & 3,011                                & 0.0103              \\ \hline
\multirow{5}{*}{\textbf{No. Steps}}                        & 0                       & 11,804                               & 0.0406              \\ \cline{2-4} 
                                                           & 1-2                     & 47,474                               & 0.1631              \\ \cline{2-4} 
                                                           & 3-6                     & 167,160                              & 0.5743              \\ \cline{2-4} 
                                                           & 7-10                    & 50,128                               & 0.1722              \\ \cline{2-4} 
                                                           & 11-21                   & 14,510                               & 0.0498              \\ \hline
\multirow{3}{*}{\textbf{Cyclomatic Complexity}}            & 1                       & 271,808                              & 0.9338              \\ \cline{2-4} 
                                                           & 2                       & 14,070                               & 0.0483              \\ \cline{2-4} 
                                                           & 3-8                     & 5,198                                & 0.0179              \\ \hline
\end{tabular}
}
\end{table}

\subsection{Prompt Engineering}
\subsubsection{System Prompt}
\label{subsubsec:sysprompt}
\tbl{tab:system_prompt} shows the full content for system prompts.

\begin{table}[h]
\caption{System Prompts for various tasks} 
\label{tab:system_prompt}
\centering
\tiny
\resizebox{\columnwidth}{!}{ 
% \begin{tabular}{>{\centering}p{0.2cm} p{6.8cm}}
\begin{tabular}{c|p{8cm}}
\hline
\textbf{Task} & \multicolumn{1}{c}{\textbf{System Prompts}} \\
\hline
\textbf{T1} & You are a software engineer. Please generate a YAML file based on the user's input below. No additional explanation is needed. The output format should be \textasciigrave \textasciigrave \textasciigrave yaml \textless Workflow\textgreater \textasciigrave \textasciigrave \textasciigrave.\\ 
\hline
\textbf{T2} & You are a software engineer. Please help the user identify whether a GitHub workflow has syntax errors or not. The output format should be ``\textless{}Yes or No\textgreater \ $\mid$ line number: \textless{}line number\textgreater{}''. If the error is in the shell script, output the line number of the ``run'' key. If no syntax errors exist, the line number is ``N/A''.  \\ 
\hline
\textbf{T3} & You are a security engineer. Please help the user detect code injection vulnerabilities in the GitHub workflow. If no vulnerabilities are detected, print ``No''. Otherwise, print ``Yes'' followed by the line number of the ``run'' key (sink in the shell script) or the line number of the ``uses'' key (sink in the GitHub Action or reusable workflow), tainted variable, and the corresponding taint source. Output format should be ``Yes $\mid$ line number: \textless{}line number\textgreater \ $\mid$ tainted variable: \textless{}tainted variable\textgreater \ $\mid$ taint source: \textless{}taint source\textgreater{}". If vulnerabilities sink inside the GitHub Action or reusable workflow, both the tainted variable and the taint source are ``N/A". \\ 
\hline
\textbf{T4} & You are a software engineer. Please help the user fix syntax errors in the GitHub workflow. The output format should be \textasciigrave \textasciigrave \textasciigrave yaml \textless Workflow\textgreater \textasciigrave \textasciigrave \textasciigrave. If the input workflow contains line numbers, please remove them from the output. \\ 
\hline
\textbf{T5} & You are a security engineer. Please help the user repair code injection vulnerabilities in the GitHub workflow. The output format should be \textasciigrave \textasciigrave \textasciigrave yaml \textless Workflow\textgreater \textasciigrave \textasciigrave \textasciigrave. If the input workflow contains line numbers, please remove them from the output.  \\ 
\hline
\end{tabular}}
\end{table}

\subsubsection{User Prompt}
\tbl{tab:prompt_eg} provides components of the user prompt for the~\gh{} workflow in~\lst{listing:wfsgen_example_gt}.
\begin{listing}[t]
\begin{minted}[breaklines, fontsize=\scriptsize]{yaml}
on:
  push:
    branches:
      - main
name: Generate Docs
jobs:
  build:
    runs-on: ubuntu-latest
    steps:
      - name: Checkout sources
        uses: actions/checkout@v2
      - name: Install stable toolchain
        uses: actions-rs/toolchain@v1
        with:
          profile: minimal
          target: wasm32-wasi
          toolchain: stable
          override: true
      - name: Run cargo build
        uses: actions-rs/cargo@v1
        with:
          command: doc
          args: --workspace --no-deps
      - name: Prepare docs folder
        run: |
          sudo chown -R $(whoami) target/doc
          touch target/doc/.nojekyll
      - name: Deploy documentation branch
        uses: JamesIves/github-pages-deploy-action@3.7.1
        with:
          GITHUB_TOKEN: ${{ secrets.GITHUB_TOKEN }}
          BRANCH: gh-pages
          FOLDER: target/doc
\end{minted}
\caption{docs.yaml in the neoeinstein/cj4-fadec\cite{egrepo} repo.}
\label{listing:wfsgen_example_gt}
\end{listing}

\begin{listing}[t]
\begin{minted}[breaklines, fontsize=\scriptsize, escapeinside=&&]{yaml}
name: CI
on:
  ...
jobs:
  test:
    name: Julia ${{ matrix.version }} - ${{ matrix.os }} - ${{ matrix.arch }}
    runs-on: ${{ matrix.os }}
    strategy:
      fail-fast: false
      matrix:
        version:
          - '1.0'
          - '1'
        os:
          - ubuntu-latest
          - macOS-latest
          - windows-latest
        arch:
          - x64
        # 32-bit Julia binaries are not available on macOS
        exclude:
          - os: macOS-latest
            arch: &\textcolor{syntxbugcolor}{\faBug}& x86 
    steps:
      ....

\end{minted}

%\begin{tcolorbox} [colback=white, top=1pt, bottom=1pt, left=2pt, right=2pt]
%\textbf{~\starchat{}:} Yes
%\end{tcolorbox}

%\begin{tcolorbox} [colback=white, top=1pt, bottom=1pt, left=2pt, right=2pt]
%\textbf{~\gpt{}:} No
%\end{tcolorbox}

\caption{CI.yml in Kolaru/FiniteDifferences.jl repo contains a syntactic error(\textcolor{syntxbugcolor}{\faBug}), which was correctly identified by~\starchat{} but missed by~\gpt{}.}
\label{listing:syntaxerror_example}
\end{listing}

\begin{listing}[th]
\begin{minted}[breaklines,escapeinside=&&,fontsize=\tiny]{yaml}
name: Tag Beta

on:
  workflow_dispatch:
  push:
    branches:
      - main

jobs:
  check:
    runs-on: ubuntu-latest

    steps:
    - name: Check for release version
      uses: actions/github-script@v5
      id: check
      with:
        script: |
          const { owner, repo } = context.repo
          const regex = /\d+\.\d+\.\d+\.\d+/g;
          &\textcolor{red}{\faBug}& if (regex.test(`${{ github.event.head_commit.message }}`)) {
            const run_id = "${{ github.run_id }}";
            await github.rest.actions.cancelWorkflowRun({ owner, repo, run_id });
          } else {
            return "build"
          }
        result-encoding: string

  tag:
    needs: check
    runs-on: ubuntu-latest

    steps:
    - uses: actions/checkout@v1

    - name: Run Luacheck
      uses: nebularg/actions-luacheck@v1
      with:
        args: --no-color -q
        annotate: warning

    - name: Set beta
      uses: onemedical/action-general-autotag@main
      with:
        GITHUB_TOKEN: ${{ secrets.REPOSITORY_ACCESS_TOKEN }}
        source_file: "Aurora.toc"
        extraction_regex: "\\s*##\\s*Version\\s*:\\s*(\\d+\\.\\d+\\.\\d+.\\d+)"
        tag_format: "{version}.${{ github.run_number }}-beta"
        tag_message: "beta"

\end{minted}
\caption{Example where fine-tuned~\gpt{} correctly identified a code injection vulnerability missed by other~\acp{LLM}. }
\label{lst:gptf_vuln_dection}
\end{listing}

\begin{table}[H]
\caption{Components of the user prompt for~\lst{listing:wfsgen_example_gt}} 
\label{tab:prompt_eg}
\centering
\tiny
\begin{tabular}{>{\centering}p{1cm} p{7cm}}
\hline

\textbf{Workflow-level Information} & Generate a GitHub Workflow named ``Generate Docs'' for a GitHub repository whose primary programming language is Rust. This workflow will be triggered by an event: The workflow would run whenever there is a push event to a branch named ``main''.\\ 
\hline
\textbf{All Job IDs} & The workflow has one job. The job id of the 1st job is ``build''.  \\ 
\hline
\textbf{Job-level Information} & The job ``build'' runs on ubuntu-latest runner. \\ 
\hline
\textbf{All Step Names} & The job ``build'' has 5 steps. The 1st step is named ``Checkout sources''. The 2nd step is named ``Install stable toolchain''. The 3rd step is named ``Run cargo build''. The 4th step is named ``Prepare docs folder''. The 5th step is named ``Deploy documentation branch''. \\ 
\hline
\textbf{Step-level Information} & The job ``build'' has 5 steps. The 1st step is named ``Checkout sources''. This step runs action ``actions/checkout'' tagged as v2. The 2nd step is named ``Install stable toolchain''. This step runs action ``actions-rs/toolchain'' tagged as v1. The step defines 4 input parameters for the action: ``profile'' is set to ``minimal'', ``target'' is set to ``wasm32-wasi'', ``toolchain'' is set to ``stable'' and ``override'' is set to ``True''. The 3rd step is named ``Run cargo build''. This step runs action ``actions-rs/cargo'' tagged as v1. The step defines 2 input parameters for the action: ``command'' is set to ``doc'' and ``args'' is set to ``{--}workspace {--}no-deps''. The 4th step is named ``Prepare docs folder''. This step runs a script: ``sudo chown -R \$(whoami) target/doc\textbackslash n touch target/doc/.nojekyll''. The 5th step is named ``Deploy documentation branch''. This step runs action ``JamesIves/github-pages-deploy-action'' tagged as 3.7.1. The step defines 3 input parameters for the action: ``GITHUB\_TOKEN'' is set to ``\$\{\{secrets.GITHUB\_TOKEN\}\}'', ``BRANCH'' is set to ``gh-pages'' and ``FOLDER'' is set to ``target/doc''. \\
\hline
\textbf{Dependencies} & Here are some Github Actions that might be used in the workflow: v2 version of actions/checkout, v1 version of actions-rs/toolchain, v1 version of actions-rs/cargo and 3.7.1 version of JamesIves/github-pages-deploy-action. \\ 
\hline
\end{tabular}
\end{table}

\subsubsection{Details in User Prompt}
\label{subsubsec:promptdetails}
\noindent \textbf{Hint message} for T3 contains a list of taint sources, and details of Actions from the Taint Summary Database~\cite{muralee2023Argus}.

\noindent \textbf{Error message} for T4 is obtained from~\actionlint{}~\cite{actionlint},~\eg{}~\textit{Invalid activity type ``started" for ``watch" Webhook event. Available types are ``starred"}. 

\noindent \textbf{Guidance on resolution} for T5: Avoid string interpolation by using an intermediate environment variable using a `env' tag to store the sensitive data, and then use the environment variable in the command under the `run' keyword by accessing it without the '\$\{\{ ... \}\}'. 

\subsection{Effectiveness of Code Injection Vulnerability Repair}
\label{apdx:codeinjecrepair}

We evaluate this task using three prompts (P1, P2, P3) with increasing detail (\tbl{tab:methodology}).
{
\renewcommand{\arraystretch}{1.1} % Increase the row spacing
\setlength{\tabcolsep}{6pt} % Increase the column spacing
\begin{table}[h]
    \centering
\resizebox{\columnwidth}{!}{% 
\begin{tabular}{|c|*{14}{c|}}
    \hline
    \multirow{3}{*}{\textbf{Model}} & \multirow{3}{*}{\textbf{t}} & \multicolumn{6}{c|}{\textbf{off-the-shelf}} & \multicolumn{6}{c|}{\textbf{fine-tuned}} \\
    \cline{3-14}
    & & \multicolumn{3}{c|}{\textbf{0-shot}} & \multicolumn{3}{c|}{\textbf{1-shot}} & \multicolumn{3}{c|}{\textbf{0-shot}} & \multicolumn{3}{c|}{\textbf{1-shot}}\\
    \cline{3-14}
    & & P1 & P2 & P3 & P1 & P2 & P3 & P1 & P2 & P3 & P1 & P2 & P3 \\ \hline
    \multirow{5}{*}{\textbf{\gpt{}}}
    & 0.1 & \shadecell{2.020} & \shadecell{2.020} & \shadecell{20.41} & \shadecell{8.050} & \shadecell{4.600} & \shadecell{31.40} & \shadecell{21.21} & \shadecell{49.49} & \shadecell{50.00} & \shadecell{26.67} & \shadecell{25.68} & \shadecell{36.99} \\
    & 0.3 & \shadecell{3.030} & \shadecell{6.060} & \shadecell{26.53} & \shadecell{12.50} & \shadecell{5.620} & \shadecell{41.86} & \shadecell{38.38} & \shadecell{52.53} & \shadecell{56.12} & \shadecell{42.67} & \shadecell{31.08} & \shadecell{49.32} \\
    & 0.5 & \shadecell{7.070} & \shadecell{9.090} & \shadecell{32.65} & \shadecell{10.11} & \shadecell{7.870} & \shadecell{47.62} & \shadecell{40.40} & \shadecell{62.63} & \shadecell{62.24} & \shadecell{42.67} & \shadecell{41.89} & \shadecell{54.79} \\
    & 0.7 & \shadecell{6.060} & \shadecell{8.080} & \shadecell{43.88} & \shadecell{13.79} & \shadecell{6.980} & \shadeunderlinecell{66.67} & \shadecell{49.49} & \shadecell{63.64} & \shadecell{69.39} & \shadecell{46.67} & \shadecell{45.95} & \shadeunderlinecell{64.38} \\
    & 0.9 & \shadecell{14.14} & \shadecell{24.24} & \shadeunderlinecell{44.90} & \shadecell{20.69} & \shadecell{15.12} & \shadecell{55.95} & \shadecell{52.53} & \shadeunderlinecell{69.70} & \shadecell{67.35} & \shadecell{53.33} & \shadecell{48.65} & \shadeunderlinecell{64.38} \\ \hline
    \multirow{5}{*}{\textbf{\codellama{}}}
    & 0.1 & \shadecell{36.00} & \shadecell{48.00} & \shadecell{26.00} & \shadecell{33.00} & \shadecell{62.00} & \shadecell{41.00} & \shadecell{0.000} & \shadecell{0.000} & \shadecell{0.000}  & \shadecell{0.000} & \shadecell{0.000} & \shadecell{1.000} \\
    & 0.3 & \shadecell{35.00} & \shadecell{52.00} & \shadecell{36.00} & \shadecell{46.00} & \shadeunderlinecell{73.00} & \shadecell{60.00} & \shadecell{0.000} & \shadecell{0.000} & \shadecell{0.000}  & \shadecell{0.000} & \shadecell{1.000} & \shadecell{1.000} \\
    & 0.5 & \shadecell{35.00} & \shadecell{42.00} & \shadecell{38.00} & \shadecell{46.00} & \shadecell{72.00} & \shadecell{67.00} & \shadecell{0.000} & \shadecell{0.000} & \shadecell{0.000}  & \shadecell{0.000} & \shadecell{1.000} & \shadecell{1.000} \\
    & 0.7 & \shadecell{32.00} & \shadecell{44.00} & \shadecell{50.00} & \shadecell{47.00} & \shadecell{59.00} & \shadecell{67.00} & \shadecell{0.000} & \shadecell{0.000} & \shadecell{0.000}  & \shadecell{0.000} & \shadecell{2.000} & \shadecell{3.000} \\
    & 0.9 & \shadecell{45.00} & \shadeunderlinecell{60.00} & \shadecell{45.00} & \shadecell{48.00} & \shadecell{57.00} & \shadecell{68.00} & \shadecell{0.000} & \shadeunderlinecell{2.000} & \shadecell{0.000}  & \shadecell{4.000} & \shadecell{4.000} & \shadeunderlinecell{7.000} \\ \hline
    \multirow{5}{*}{\textbf{\starchat{}}}
    & 0.1 & \shadecell{1.000} & \shadecell{4.000} & \shadecell{11.00} & \shadecell{2.000} & \shadecell{2.000} & \shadecell{3.000} & \shadecell{0.000} & \shadecell{0.000} & \shadecell{0.000}  & \shadecell{0.000} & \shadecell{0.000} & \shadecell{0.000} \\
    & 0.3 & \shadecell{2.000} & \shadecell{8.000} & \shadecell{17.00} & \shadecell{3.000} & \shadecell{3.000} & \shadecell{5.000} & \shadecell{0.000} & \shadecell{0.000} & \shadecell{0.000}  & \shadecell{0.000} & \shadecell{0.000} & \shadecell{0.000} \\
    & 0.5 & \shadecell{7.000} & \shadecell{10.00} & \shadecell{20.00} & \shadecell{3.000} & \shadecell{6.000} & \shadecell{11.00} & \shadecell{0.000} & \shadecell{0.000} & \shadecell{0.000}  & \shadecell{0.000} & \shadecell{0.000} & \shadecell{0.000} \\
    & 0.7 & \shadecell{9.000} & \shadecell{20.00} & \shadecell{29.00} & \shadecell{7.000} & \shadecell{5.000} & \shadecell{13.00} & \shadecell{0.000} & \shadecell{0.000} & \shadecell{0.000}  & \shadecell{0.000} & \shadecell{0.000} & \shadecell{0.000} \\
    & 0.9 & \shadecell{21.00} & \shadecell{19.00} & \shadeunderlinecell{37.00} & \shadecell{10.00} & \shadecell{10.00} & \shadeunderlinecell{19.00} & \shadecell{0.000} & \shadecell{0.000} & \shadecell{0.000}  & \shadeunderlinecell{2.000} & \shadecell{0.000} & \shadecell{0.000} \\ \hline
    \end{tabular}
    
}
    \caption{Accuracy@K of code injection vulnerability repair on~\ciset{}.}
    \label{tab:wfsvulfix_step1_argus_acc}
\end{table}
}

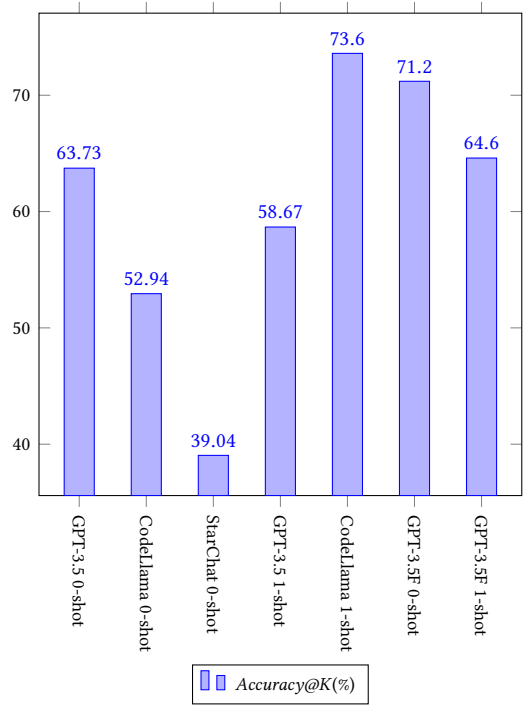
\begin{figure}[h]
    \centering
      % \begin{adjustbox}{right=8cm}    
    \begin{tikzpicture}
    \begin{axis}[
        ybar,
        width=8cm,
        height=8cm,
        bar width=.4cm,
        legend style={at={(0.5,-0.35)},anchor=north,legend columns=-1},
        symbolic x coords={GPT-3.5 0-shot,CodeLlama 0-shot,StarChat 0-shot,GPT-3.5 1-shot,CodeLlama 1-shot,GPT-3.5F 0-shot,GPT-3.5F 1-shot},
        xtick=data,
        xticklabel style={rotate=270},
        nodes near coords,
        every node near coord/.append style={font=\small},
        ] 
    \addplot+ coordinates {(GPT-3.5 0-shot,63.73) (CodeLlama 0-shot,52.94) (StarChat 0-shot,39.04) (GPT-3.5 1-shot,58.67) (CodeLlama 1-shot,73.60) (GPT-3.5F 0-shot,71.20) (GPT-3.5F 1-shot,64.60)};
    
    \legend{~\acck{}(\%)}
    \end{axis}
    \end{tikzpicture}
    % \end{adjustbox}
    \caption{Final evaluation for code injection vulnerability repair.}
    
    \label{fig:wfsvulfix_step2}
\end{figure}

\noindent{\emph{Calibration:}} The~\tbl{tab:wfsvulfix_step1_argus_acc} shows the~\acck{} of different~\acp{LLM} on our calibration dataset (\ciset{}).
Similar to syntactic error repair, higher temperatures yield better results across all prompts.
Also, detailed prompts provide better results, as indicated by the increasing trend across P1 to P3.
Fine-tuned variant of~\gpt{} performs better on code injection vulnerability tasks (unseen tasks) than the off-the-shelf variant.
However, the case is different with~\codellama{} and~\starchat{}, where the fine-tuned variant performs poorly.
These results demonstrate that fine-tuning~\gpt{} on certain tasks also helps in improving its effectiveness on other unseen but related tasks.
However, this is not the case with other~\acp{LLM}, where fine-tuned variants can perform poorly on unseen (but related) tasks.

\noindent{\emph{Final Evaluation:}}
\fig{fig:wfsvulfix_step2} shows the evaluation on the final large dataset. We did not include the results for \starchat{} in 1-shot mode and fine-tuned variants of \codellama{} and \starchat{} since they are extremely poor.~\codellama{} in 1-shot mode performs the best across all LLMs and their variants.

\subsection{Size v/s Workflow Generation Effectiveness}
\label{apdx:sizeworkflowgeneration}
\begin{figure}[H]
    \pgfplotsset{footnotesize}
    \caption{~\bleu{} score (left) and~\acck{} (right) against the size (in KB) of expected workflows.}
    \begin{center}
    \begin{tikzpicture}
        \begin{axis}
        \addplot [blue, smooth, mark=triangle] table[x=x, y=y] {data/bleu_size/gpt-3.5_0shot.dat};
        \addplot [red, smooth, mark=triangle] table[x=x, y=y] {data/bleu_size/gpt-3.5_1shot.dat};
        \addplot [green, smooth, mark=triangle] table[x=x, y=y] {data/bleu_size/finetuned-gpt-3.5_0shot.dat};

        \addplot [blue, smooth, mark=*] table[x=x, y=y] {data/bleu_size/starchat_0shot.dat};
        \addplot [red, smooth, mark=*] table[x=x, y=y] {data/bleu_size/starchat_1shot.dat};
        \addplot [green, smooth, mark=*] table[x=x, y=y] {data/bleu_size/starchat-finetuned_0shot.dat};

        \addplot [blue, smooth, mark=+] table[x=x, y=y] {data/bleu_size/codellama-instruct_0shot.dat};
        \addplot [red, smooth, mark=+] table[x=x, y=y] {data/bleu_size/codellama-instruct_1shot.dat};
        \addplot [green, smooth, mark=+] table[x=x, y=y] {data/bleu_size/codellama-instruct-finetuned_0shot.dat};
        \end{axis}
    \end{tikzpicture}
    \begin{tikzpicture}
        \begin{axis}[
            legend columns=3,
            legend to name=named,
        ]

        \addplot [blue, smooth, mark=triangle] table[x=x, y=y] {data/acc_size/gpt-3.5_0shot.dat};
        \addlegendentry{GPT-3.5 0-shot}

        \addplot [red, smooth, mark=triangle] table[x=x, y=y] {data/acc_size/gpt-3.5_1shot.dat};
        \addlegendentry{GPT-3.5 1-shot}

        \addplot [green, smooth, mark=triangle] table[x=x, y=y] {data/acc_size/finetuned-gpt-3.5_0shot.dat};
        \addlegendentry{GPT-3.5F}

        \addplot [blue, smooth, mark=*] table[x=x, y=y] {data/acc_size/starchat_0shot.dat};
        \addlegendentry{StarChat 0-shot}

        \addplot [red, smooth, mark=*] table[x=x, y=y] {data/acc_size/starchat_1shot.dat};
        \addlegendentry{StarChat 1-shot}

        \addplot [green, smooth, mark=*] table[x=x, y=y] {data/acc_size/starchat-finetuned_0shot.dat};
        \addlegendentry{StarChatF}

        \addplot [blue, smooth, mark=+] table[x=x, y=y] {data/acc_size/codellama-instruct_0shot.dat};
        \addlegendentry{CodeLlama 0-shot}

        \addplot [red, smooth, mark=+] table[x=x, y=y] {data/acc_size/codellama-instruct_1shot.dat};
        \addlegendentry{CodeLlama 1-shot}

        \addplot [green, smooth, mark=+] table[x=x, y=y] {data/acc_size/codellama-instruct-finetuned_0shot.dat};
        \addlegendentry{CodeLlamaF}

        \end{axis}
        \end{tikzpicture}
    \ref{named}
     \end{center}
    \label{fig:bleu_acc_size}
\end{figure}
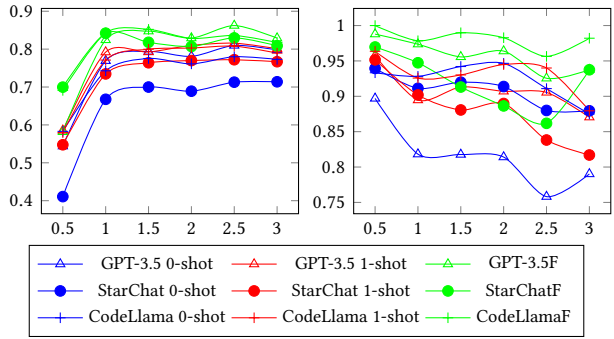
The left subfigure of~\fig{fig:bleu_acc_size} shows the trend of~\bleu{} scores against the size (in KB) of expected workflows. 
As we show in~\tbl{tab:dataset}, most of the workflows are less than 4KB. Specifically, the majority (> 85\%) of workflows are less than 3KB.
For our size-related comparisons (\ie{}~\fig {fig:bleu_acc_size}), we only considered workflows up to 3KB.
We can see from~\fig{fig:bleu_acc_size} (left subfigure) that as the workflow size increases,~\bleu{} score initially increases rapidly and then remains unchanged.

As shown by the right subfigure of~\fig{fig:bleu_acc_size}, unlike~\bleu{} scores, the~\acck{} scores slowly decrease as workflow size becomes larger.
This is expected as workflow size increases,~\acp{LLM} need to generate more tokens, increasing the likelihood of generating syntax errors resulting in lower~\acck{} scores.

\subsection{Size v/s Defect Detection Effectiveness}
\label{apdx:sizedefectdetection}
We know that the majority of workflows are less than 3KB from \apdx{apdx:sizeworkflowgeneration}, and so we only consider workflows less than 3 KB here.
The \fscore{} trend remains consistent across workflow sizes for both the identification of syntactic errors and the detection of code injection vulnerabilities. 
Consequently, we will not include a figure illustrating the \fscore{} trend relative to workflow size.

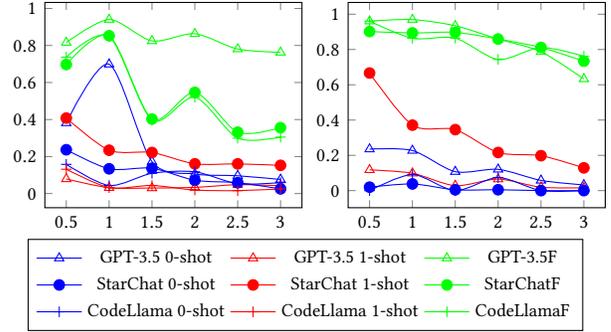
\begin{figure}[t]
    \pgfplotsset{footnotesize}
    \caption{~\acck{} of syntactic error identification (left) and code injection vulnerability detection (right) against the size (in KB) of the workflows.}
    \begin{center}
    \begin{tikzpicture}
        \begin{axis}[
            legend columns=3,
            legend to name=named,
        ]
        \addplot [blue, smooth, mark=triangle] table[x=x, y=y] {data/syntax_error_detection_acc-size/gpt-3.5_0shot.dat};
        \addlegendentry{GPT-3.5 0-shot}
        
        \addplot [red, smooth, mark=triangle] table[x=x, y=y] {data/syntax_error_detection_acc-size/gpt-3.5_1shot.dat};
        \addlegendentry{GPT-3.5 1-shot}
        
        \addplot [green, smooth, mark=triangle] table[x=x, y=y] {data/syntax_error_detection_acc-size/finetuned-gpt-3.5_0shot.dat};
        \addlegendentry{GPT-3.5F}
        
        \addplot [blue, smooth, mark=*] table[x=x, y=y] {data/syntax_error_detection_acc-size/starchat_0shot.dat};
        \addlegendentry{StarChat 0-shot}
        
        \addplot [red, smooth, mark=*] table[x=x, y=y] {data/syntax_error_detection_acc-size/starchat_1shot.dat};
        \addlegendentry{StarChat 1-shot}
        
        \addplot [green, smooth, mark=*] table[x=x, y=y] {data/syntax_error_detection_acc-size/starchat-finetuned_0shot.dat};
        \addlegendentry{StarChatF}

        \addplot [blue, smooth, mark=+] table[x=x, y=y] {data/syntax_error_detection_acc-size/codellama-instruct_0shot.dat};
        \addlegendentry{CodeLlama 0-shot}

        \addplot [red, smooth, mark=+] table[x=x, y=y] {data/syntax_error_detection_acc-size/codellama-instruct_1shot.dat};
        \addlegendentry{CodeLlama 1-shot}

        \addplot [green, smooth, mark=+] table[x=x, y=y] {data/syntax_error_detection_acc-size/codellama-instruct-finetuned_0shot.dat};
        \addlegendentry{CodeLlamaF}

        \end{axis}
    \end{tikzpicture}
    \begin{tikzpicture}
        \begin{axis}

        \addplot [blue, smooth, mark=triangle] table[x=x, y=y] {data/vuln_detection_acc-size/gpt-3.5_0shot.dat};

        \addplot [red, smooth, mark=triangle] table[x=x, y=y] {data/vuln_detection_acc-size/gpt-3.5_1shot.dat};

        \addplot [green, smooth, mark=triangle] table[x=x, y=y] {data/vuln_detection_acc-size/finetuned-gpt-3.5_0shot.dat};

        \addplot [blue, smooth, mark=*] table[x=x, y=y] {data/vuln_detection_acc-size/starchat_0shot.dat};

        \addplot [red, smooth, mark=*] table[x=x, y=y] {data/vuln_detection_acc-size/starchat_1shot.dat};

        \addplot [green, smooth, mark=*] table[x=x, y=y] {data/vuln_detection_acc-size/starchat-finetuned_0shot.dat};

        \addplot [blue, smooth, mark=+] table[x=x, y=y] {data/vuln_detection_acc-size/codellama-instruct_0shot.dat};

        \addplot [green, smooth, mark=+] table[x=x, y=y] {data/vuln_detection_acc-size/codellama-instruct-finetuned_0shot.dat};

        \end{axis}
        \end{tikzpicture}
    \ref{named}
     \end{center}
    \label{fig:acc-size}
\end{figure}

\fig{fig:acc-size} shows the trend of~\acck{} against the size (in KB) of the workflows for syntactic error identification and code injection vulnerability detection. 
For syntactic error identification, \gpt{} in 0-shot mode and the fine-tuned variants of \starchat{} and \codellama{}
have considerable fluctuations. Other models remain constant as the size of the workflow increases.
For code injection vulnerability detection, apart from \starchat{} in 1-shot mode, which exhibits a significant decrease in \acck{} as the workflow size increases, the \acck{} of other models remains relatively unchanged.

\subsection{Figures and Tables}

{
\renewcommand{\arraystretch}{1.1} % Increase the row spacing
\setlength{\tabcolsep}{6pt} % Increase the column spacing
\begin{table}[h]
    \caption{Accuracy@K of code-injection vulnerability detection on~\ciset{}.}
    \vspace{-9pt}

    \centering
 \resizebox{\columnwidth}{!}{% 
    \small
    \begin{tabular}{|c|*{11}{c|}}
    \hline
    \multirow{3}{*}{\textbf{Model}} & \multirow{3}{*}{\textbf{t}} & \multicolumn{6}{c|}{\textbf{off-the-shelf}} & \multicolumn{3}{c|}{\multirow{2}{*}{\textbf{fine-tuned}}} \\
    \cline{3-8}
    & & \multicolumn{3}{c|}{\textbf{0-shot}} & \multicolumn{3}{c|}{\textbf{1-shot}} & \multicolumn{3}{c|}{}\\
    \cline{3-11}
    & & P1 & P2 & P3 & P1 & P2 & P3 & P1 & P2 & P3 \\ \hline
    \multirow{5}{*}{\textbf{\gpt{}}}
    & 0.1 & \shadecell{0.000} & \shadecell{0.000} & \shadecell{8.570} & \shadecell{0.000} & \shadecell{0.000} & \shadecell{3.850} & \shadecell{86.67} & \shadecell{89.52} & \shadecell{89.52} \\
    & 0.3 & \shadecell{0.000} & \shadecell{0.000} & \shadecell{9.520} & \shadecell{0.950} & \shadecell{0.000} & \shadecell{9.620} & \shadecell{86.67} & \shadecell{90.48} & \shadecell{90.48} \\
    & 0.5 & \shadecell{0.000} & \shadecell{0.000} & \shadecell{12.38} & \shadecell{0.950} & \shadecell{0.000} & \shadeunderlinecell{12.50} & \shadecell{89.52} & \shadecell{91.43} & \shadecell{91.43} \\
    & 0.7 & \shadecell{0.000} & \shadecell{0.000} & \shadecell{9.520} & \shadecell{0.000} & \shadecell{0.000} & \shadeunderlinecell{12.50} & \shadecell{88.57} & \shadecell{93.33} & \shadecell{89.52} \\
    & 0.9 & \shadecell{0.000} & \shadecell{0.950} & \shadeunderlinecell{12.38} & \shadecell{0.950} & \shadecell{0.000} & \shadeunderlinecell{12.50} & \shadecell{90.48} & \shadeunderlinecell{96.19} & \shadecell{90.48} \\ \hline
    \multirow{5}{*}{\textbf{\codellama{}}}
    & 0.1 & \shadecell{0.950} & \shadecell{0.950} & \shadeunderlinecell{3.810} & \shadecell{0.000} & \shadecell{0.000} & \shadecell{0.000} & \shadecell{60.00} & \shadecell{55.24} & \shadecell{53.33} \\
    & 0.3 & \shadecell{0.950} & \shadecell{0.000} & \shadecell{2.860} & \shadecell{0.000} & \shadecell{0.000} & \shadecell{0.000} & \shadecell{64.76} & \shadecell{66.67} & \shadecell{58.10} \\
    & 0.5 & \shadecell{0.000} & \shadecell{0.000} & \shadeunderlinecell{3.810} & \shadecell{0.000} & \shadecell{0.000} & \shadecell{0.000} & \shadecell{65.71} & \shadecell{69.52} & \shadecell{60.95} \\
    & 0.7 & \shadecell{0.950} & \shadecell{0.000} & \shadecell{1.900} & \shadecell{0.000} & \shadecell{0.000} & \shadecell{0.000} & \shadecell{67.62} & \shadeunderlinecell{71.43} & \shadecell{62.86} \\
    & 0.9 & \shadecell{0.000} & \shadecell{0.950} & \shadecell{0.000} & \shadecell{0.000} & \shadecell{0.000} & \shadecell{0.000} & \shadecell{63.81} & \shadecell{68.57} & \shadecell{66.67} \\ \hline
    \multirow{5}{*}{\textbf{\starchat{}}}
    & 0.1 & \shadecell{0.000} & \shadecell{0.000} & \shadecell{1.900} & \shadecell{12.38} & \shadecell{17.14} & \shadecell{20.95} & \shadecell{68.57} & \shadecell{68.57} & \shadecell{64.76} \\
    & 0.3 & \shadecell{0.000} & \shadecell{0.000} & \shadecell{0.950} & \shadecell{18.10} & \shadecell{19.05} & \shadecell{25.71} & \shadecell{74.29} & \shadecell{76.19} & \shadecell{65.71} \\
    & 0.5 & \shadecell{0.000} & \shadecell{0.000} & \shadeunderlinecell{2.860} & \shadecell{12.38} & \shadecell{22.86} & \shadeunderlinecell{28.57} & \shadecell{78.10} & \shadecell{75.24} & \shadecell{68.57} \\
    & 0.7 & \shadecell{0.000} & \shadecell{0.000} & \shadeunderlinecell{2.860} & \shadecell{12.38} & \shadecell{16.19} & \shadecell{25.71} & \shadecell{73.33} & \shadecell{77.14} & \shadecell{72.38} \\
    & 0.9 & \shadecell{0.000} & \shadecell{0.000} & \shadecell{1.900} & \shadecell{6.670} & \shadecell{10.48} & \shadecell{26.67} & \shadeunderlinecell{80.95} & \shadecell{78.10} & \shadecell{74.29} \\ \hline
    \end{tabular}
        }
    \label{tab:wfsvulfind_step1_acc}
\end{table}
}
\tbl{tab:wfsvulfind_step1_acc} shows the~\acck{} of code injection vulnerability detection of different models and their variants across different modes on~\ciset{}.

\begin{figure}[H]

    \begin{adjustbox}{right=8cm}
    \begin{tikzpicture}
    \begin{axis}[
        xbar,
        xmin=0,
        xmax=110,
        grid=major,
        major grid style={dashed},
        bar width=.2cm,
        width=8cm,
        height=8cm,
        legend style={at={(0.5,-0.1)},anchor=north,legend columns=-1},
        symbolic y coords={GPT-3.5 0-shot,CodeLlama 0-shot,StarChat 0-shot,GPT-3.5 1-shot,CodeLlama 1-shot,StarChat 1-shot,GPT-3.5F,CodeLlamaF,StarChatF},
        ytick=data,
        yticklabel style={text width=1.5cm,align=right},
        nodes near coords,
        nodes near coords align={horizontal},
        every node near coord/.append style={font=\small},
        ]
    \draw [purple!20, fill] (axis description cs:0, 0.66) rectangle (axis description cs:1,0.97); 
    \draw [orange!20, fill] (axis description cs:0, 0.35) rectangle (axis description cs:1,0.65); 
    \draw [yellow!20, fill] (axis description cs:0, 0.03) rectangle (axis description cs:1,0.34); 
    \addplot+ coordinates {(66.18,GPT-3.5 0-shot) (34.70,CodeLlama 0-shot) (66.16,StarChat 0-shot) (2.41,GPT-3.5 1-shot) (23.24,CodeLlama 1-shot) (100.0,StarChat 1-shot) (91.88,GPT-3.5F) (81.18,CodeLlamaF) (82.67,StarChatF)};
    \addplot+ coordinates {(41.36,GPT-3.5 0-shot) (7.04,CodeLlama 0-shot) (11.28,StarChat 0-shot) (3.64,GPT-3.5 1-shot) (3.2,CodeLlama 1-shot) (21.04,StarChat 1-shot) (87.80,GPT-3.5F) (63.24,CodeLlamaF) (64.55,StarChatF)};
    
    \legend{~\fscore{}(\%), ~\acck{}(\%)}
    \end{axis}
    \end{tikzpicture}
    \end{adjustbox}
        \caption{Final evaluation for syntax error detection}
    \label{fig:wfsbugfind}
\end{figure}
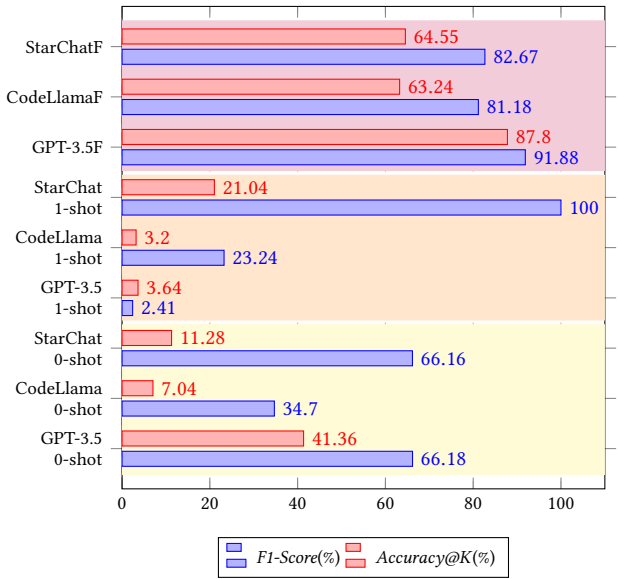

\fig{fig:wfsbugfind} shows the final evaluation of syntactic error detection of different models with the best-performing configuration on the final large dataset.

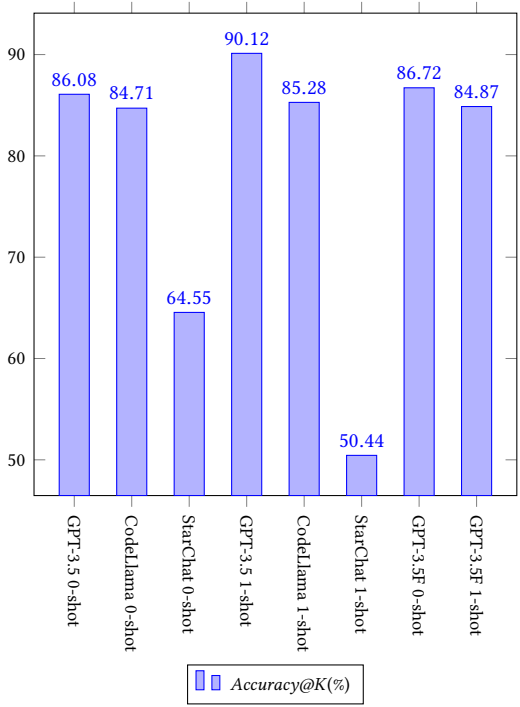
\begin{figure}[b]

    \centering
    \begin{tikzpicture}
    \begin{axis}[
        ybar,
        width=8cm,
        height=8cm,
        bar width=.4cm,
        legend style={at={(0.5,-0.35)},anchor=north,legend columns=-1},
        symbolic x coords={GPT-3.5 0-shot,CodeLlama 0-shot,StarChat 0-shot,GPT-3.5 1-shot,CodeLlama 1-shot,StarChat 1-shot,GPT-3.5F 0-shot,GPT-3.5F 1-shot},
        xtick=data,
        xticklabel style={rotate=270},
        nodes near coords,
        every node near coord/.append style={font=\small},
        ] 
    \addplot+ coordinates {(GPT-3.5 0-shot,86.08) (CodeLlama 0-shot,84.71) (StarChat 0-shot,64.55) (GPT-3.5 1-shot,90.12) (CodeLlama 1-shot,85.28) (StarChat 1-shot,50.44) (GPT-3.5F 0-shot,86.72) (GPT-3.5F 1-shot,84.87)};
    
    \legend{~\acck{}(\%)}
    \end{axis}
    \end{tikzpicture}
        \caption{Final evaluation for syntactic error repair.}
    \label{fig:wfsbugfix_step2}
\end{figure}

\fig{fig:wfsbugfix_step2} shows $Accuracy@k$ of syntactic error fixing on the large dataset.

\end{document}